\documentclass[11pt,a4paper]{article}
 \pdfoutput=1
\usepackage{jcappub}

\usepackage{tabularx}
\usepackage{subcaption}
\usepackage{caption}
\usepackage{enumerate}
\usepackage{amsmath,amssymb}
\usepackage{graphicx}
\usepackage{dcolumn}
\usepackage{bm,color}
\usepackage{hyperref}
\usepackage{accents}
\usepackage{amssymb,float}
\usepackage{amsmath}
\usepackage{multirow}
\usepackage{pifont}
\newcommand{\cmark}{\ding{51}}%
\newcommand{\xmark}{\ding{55}}%

\hypersetup{
    colorlinks=true,
    linkcolor=red,
    filecolor=magenta,      
    citecolor=blue
}

\title{\boldmath Performance of Non-Parametric Reconstruction Techniques in the Late-Time Universe}

\author[a]{Celia Escamilla-Rivera,}
\author[b,c]{Jackson Levi Said}
\author[b]{and Jurgen Mifsud}

\affiliation[a]{Instituto de Ciencias Nucleares, Universidad Nacional Aut\'{o}noma de M\'{e}xico, Circuito Exterior C.U., A.P. 70-543, M\'exico D.F. 04510, M\'{e}xico.}
\affiliation[b]{Institute of Space Sciences and Astronomy, University of Malta, Malta, MSD 2080}
\affiliation[c]{Department of Physics, University of Malta, Malta, MSD 2080}

\emailAdd{celia.escamilla@nucleares.unam.mx}
\emailAdd{jackson.said@um.edu.mt}
\emailAdd{jurgen.mifsud@um.edu.mt}

\abstract{
In the context of a Hubble tension problem that is growing in its statistical significance, we reconsider the effectiveness of non-parametric reconstruction techniques which are independent of prescriptive cosmological models. By taking cosmic chronometers, Type Ia Supernovae and baryonic acoustic oscillation data, we compare and contrast two important reconstruction approaches, namely Gaussian processes (GP) and the \textbf{Lo}cally w\textbf{e}ighted \textbf{S}catterplot \textbf{S}moothing together with \textbf{Sim}ulation and \textbf{ex}trapolation method (LOESS-Simex or LS). In the context of these methods, besides not requiring a cosmological model, they also do not require physical parameters in their approach to their reconstruction of data (but they do depend on statistical hyperparameters). We firstly show how both GP and LOESS-Simex can be used to successively reconstruct various data sets to a high level of precision. We then directly compare both approaches in a quantitative manner by considering several factors, such as how well the reconstructions approximate the data sets themselves to how their respective uncertainties evolve. In light of the puzzling Hubble tension, it is important to consider how the uncertain regions evolve over redshift and the methods compare for estimating cosmological parameters at current times. For cosmic chronometers and baryonic acoustic oscillation compiled data sets, we find that GP generically produce smaller variances for the reconstructed data with a minimum value of $\sigma_{\rm GP-min} = 1.1$, while the situation for LS is totally different with a minimum of $\sigma_{\rm LS-min} = 50.8$.
Moreover, some of these characteristics can be alleviate at low $z$, where LS presents less underestimation in comparison to GP.
}

\begin{document}

\maketitle
\flushbottom

\section{Introduction}

The precise value of late time cosmic acceleration remains a great mystery that is causing serious tensions with predictions from the early Universe assuming vanilla $\Lambda$CDM \cite{Bernal:2016gxb}. Since the discovery of the accelerating expansion of the Universe in the late 90s \cite{Perlmutter:1998np,Riess:1998cb}, flat $\Lambda$CDM has emerged as the concordance model despite several theoretical challenges in the model such as 
fine-tuning issues, the horizon and coincidence problems, the cosmological constant and dark matter problems, among many more \cite{RevModPhys.61.1,Bull:2015stt}. In the background of these challenges, there have been several compelling potential solution proposals ranging from beyond $\Lambda$CDM such as dynamical dark energy \cite{Copeland:2006wr}, extended gravity \cite{Capozziello:2011et} to beyond general relativity (GR) \cite{Clifton:2011jh}, and others.

The increasing separation in $H_0$ values between early Universe $\Lambda$CDM-based predictions and local cosmology-independent measurements has become a serious threat that may soon pass over to other cosmological parameters such as the value of $f\sigma_8$ \cite{DiValentino:2020vvd,Aghanim:2018eyx}, calling into question the predictive power of $\Lambda$CDM. In this context, it is extremely important to have a range of reliable techniques to reconstruct combined data sets that can accurately determine late time cosmological parameter values \cite{10.5555/1162254,10.1145/3390557.3394126}. Given a cosmological model, the well known Markov chain Monte Carlo (MCMC) approach readily determines best-fit model parameters for cosmological data \cite{alma991007710432905601}. In recent years, interest has grown in non-parametric models which do not assume a cosmological model at all \cite{kiusalaas_2013}. These methods have largely relied on supervised learning in which the relations between the points in a data set are used to either reduce noise in the data or to simulate other points not in the observations. 

In this work, we develop comparison techniques for two popular approaches that have gained traction in recent years, namely Gaussian process (GP) regression \cite{Banerjee} , and \textbf{Lo}cally w\textbf{e}ighted \textbf{S}catterplot \textbf{S}moothing (LOESS or LS)  \cite{Cleveland}. On the one hand, GP offers an interesting avenue to constrain a covariance function which can then be used to simulate new points as well as to smoothen places where data points exist. The effect of having many points in a data set helps reduce noise in observations. The drawback of this method is that if the behavior or density of points in one region of a data set is drastically different to another then one of these areas may be poorly reconstructed. GP has been used widely in cosmology from smoothening Hubble data \cite{Gomez-Valent:2018hwc,Colgain:2021ngq,Yennapureddy:2017vvb,Li:2019nux,Seikel:2013fda,Seikel2012,Seikel:2013fda}, $f\sigma_8$ data \cite{Benisty:2020kdt}, and gravitational wave analysis \cite{Belgacem:2019zzu,Moore:2015sza,Canas-Herrera:2021qxs}. A recent development that has garnered momentum in recent years is that of using GP to reconstruct various possible extended gravity scenarios in which a precise Lagrangian function can be left relatively arbitrary. In Ref.~\cite{Reyes:2021owe} viable Horndeski gravity was shown to adequately reproduce late time Hubble expansion, while in Refs.~\cite{Briffa:2020qli,Cai:2019bdh} the inverse problem in $f(T)$ gravity was approached using Hubble data, this was then extended in Ref.~\cite{LeviSaid:2021yat} to growth data. Finally, in Ref.~\cite{Cai:2015zoa} some possible interactions between dark energy and dark matter were explored.

On the other hand, the LOESS reconstruction method is a robust and computationally non-parametric regression approach to reconstructing data. The main purpose of this technique is to avoid assuming any prior or cosmological model. In this line of thought, LOESS recovers the global trend of a data set by studying the suitable number of neighbouring data points around the pivot point. Furthermore, in order to compute the measurement errors of the data we use \textbf{Sim}ulation and \textbf{ex}trapolation method (Simex). The merging of both techniques is the so-called LOESS-Simex reconstruction, which has been used in order to reconstruct the cosmic expansion \cite{Montiel:2014fpa}, the Om diagnostic up to first derivative on the Hubble parameter \cite{Escamilla-Rivera:2015odt} and to reconstruct rotation curves for dark matter profiles \cite{Fernandez-Hernandez:2018yao}. Both the LOESS-Simex and GP approaches to reconstruction are model-independent in that they do not depend on a physical model of cosmology in order to produce reconstructions of cosmological parameters. However, they are more than this in that they do not require physical parameters beyond the one they are reconstructing, which is a significant property for these methods. On the other hand, they do require the use of statistical hyperparameters in order to function. However, these hyperparameters are used for statistical purposes and do not have a direct relationship to the physical parameters of a particular cosmological model.

\textbf{
\begin{table*}
\centering
\resizebox{\columnwidth}{!}{%
\begin{tabular}
{|p{4cm}||p{3cm}|p{3cm}|p{3cm}|p{3cm}|p{3cm}||}
 \hline
 \multicolumn{6}{|c|}{Nonparametric and model independent reconstructions approaches} \\
 \hline
 Method 
 & Assumption of prior/models 
 & Binned 
 & Low efficiency at high $z$
 & Underestimation of errors $z$
 & High computational cost
 \\
 \hline
Principal Components Analysis (PCA) \cite{Huterer:2002hy}   
& \cmark & \xmark &  \cmark  &  \cmark &   \cmark \\
Nonlinear Inverse Approach (NIA) \cite{Espana-Bonet:2005wkl}   
& \cmark & \xmark &  \cmark  &  \xmark &   \cmark \\
Dipole of the Luminosity Distance method (DLD) \cite{Bonvin:2006en} 
& \xmark & \xmark &  \cmark  &  \xmark &   \xmark \\
Nodal Reconstruction (NR) \cite{AlbertoVazquez:2012ofj}   
& \cmark & \xmark &  \xmark  &  \xmark &   \xmark \\
 Genetic Algorithms (GA)  \cite{Bogdanos_2009}   
& \xmark & \xmark &  \xmark  &  \cmark &   \xmark \\
Reconstructions of the Expansion History (MIR-I,II,III) \cite{Daly_2003,Fay_2006,Benitez_Herrera_2011}  
& \xmark & \xmark &  \xmark  &  \cmark &   \cmark \\
Gaussian Processes (GP) \cite{Holsclaw_2010} 
& \cmark & \xmark &  \xmark  &  \cmark &   \cmark \\
 Loess-Simex \cite{Montiel_2014} 
& \xmark & \xmark &  \xmark  &  \xmark &   \xmark \\
 \hline
\end{tabular}
}
\caption{A table of nonparametric and model independent reconstructions approaches. 
\textit{First column:} Method and its reference. \textit{Second column:} Indicates when a model requires an assumption of a prior ir a fiducial cosmological model, usually this leads to biased results. \textit{Third column:} Indicates when the method share the same data with the bins selected. This causes fluctuations. \textit{Fourth column:} This can leads to numerical instability. \textit{Fifth column:} Methods that require implementation of  tools to propagate errors. \textit{Sixth column:} High computational cost mostly due the increase of the data sample.}\label{table_methods}
\end{table*}
}

The literature contains several other approaches to reconstructing observational data using model-independent approaches \cite{Huterer:2002hy,Espana-Bonet:2005wkl,Bonvin:2006en,AlbertoVazquez:2012ofj,Bogdanos_2009,Daly_2003,Holsclaw_2010,Montiel_2014} although many of them are aimed at continuing data in regions of scarcity of data. To this end, we list many of the popular ones in the literature in Table \ref{table_methods} where we emphasize how they cope with error estimation, their computational cost, as well as whether they make prior assumptions on the behaviour of the data or whether they use binning to analyse the data. There is also a growing tendency to use a combination of these approaches to better reconstruct data sets such as using genetic algorithms with GP to better motivate the kernel selection problem in this method \cite{Bernardo:2021mfs}. There are also novel approaches being developed such as using artificial neural networks to reconstruct cosmological data. In this work, our aim is to compare two of the most promising and used methods in the literature in the cosmological context, and to show how they compare against each other.

In the context of cosmology and specifically in terms of Hubble data sets, these methods have not been directly compared against each other. However, more generally, reconstruction methods are not generally compared against each other which leaves open the question of which of the vast array of non-parametric approaches reproduce the data sets best? In this work, we consider both GP and LOESS-Simex reconstruction techniques in Sec.~\ref{sec:approaches} which we then apply to Hubble data in Sec.~\ref{sec:results}. Here, we also discuss how we use the observational data sets in the context of priors. In Sec.~\ref{sec:comparative}, we explore several ways of directly comparing both reconstruction approaches against each other. Finally, we discuss and summarise our core results in Sec.~\ref{sec:conclusions}.


\section{Model independent approaches to reconstructing Hubble observations}
\label{sec:approaches}

Non-parametric approaches to reconstructing observational data have been exhaustively explored in light of the possible tensions in observations and deviations in model predictions with recent precision tests, such as the $H_0$ tension \cite{Guo:2018ans,Bernal:2016gxb} as well as the growing $\sigma_{8,0}$ tension \cite{DiValentino:2020vvd,Lambiase:2018ows}. We introduce here the two methods that we will then use to compare and contrast expansion rate reconstructions that follow in the ensuing sections.


\subsection{LOESS-Simex reconstruction for Cosmology}\label{sec:Loess_intro}

LOESS is a model independent non-parametric regression method. For instance, with this method it is not necessarily to assume a prior nor do we assume anything about the cosmological model. The fact that it is local is because we use the local neighbourhood of each pivot point in turn to infer the global trend of data set at hand.

Our goal is to reconstruct the Hubble function which can explain the behaviour of the data set using LOESS. A recipe to generate the reconstructed values with this method is as follows \cite{Montiel:2014fpa}:
\begin{enumerate}
\item We choose a subset with $n$-points out of the $N$ data points in the neighbourhood of our pivot point $z_{i,0}$. The difference between the pivot point $z_{i,0}$ and the afar point of the neighbourhood ($\text{max}|z_{i,0}-z_j|$) where($j = 1,2,\ldots n$), is known as the \textit{span} or \textit{bandwidth} $h$ of this subset. As a first step we have to select the neighbouring points  around the pivot point. This can be fixed by deciding the value of the the smoothing parameter $\alpha$, which relates the optimal number of neighbourhood points ($n$) to the total number of data points ($N$), i.e. $n= \alpha N$, where $\alpha$ ranges from $0$ to $1$.
The convenient path to select the optimum value for the smoothing parameter is via a \textit{cross validation technique}.

The reconstructed value of $\eta^{obs}_i$ is obtained by removing the $i^{th}$ observation for a given value of $\alpha$. The reconstructed value of $\eta^{obs}$ is  denoted by $\widehat{\eta}_{-i}$ and this process is repeated for the whole data set.
The cross validation function (CV) is given by
\begin{equation}\label{eq:CV}
    \text{CV}(\alpha)= \frac{1}{N}\sum_{i=1}^{N}(\eta^{\text{obs}}_i-{\widehat \eta _{-i}})^2\,,
\end{equation}
where CV($\alpha$) denotes a mean squared error for different values of $\alpha$. We then compute  CV$(\alpha)$ for several values of $\alpha$. The value of $\alpha$ for which CV$(\alpha)$ is a minimum is chosen to be the optimal choice of $\alpha$ and used as the smoothing parameter for the subsequent calculations.

\item Next, we consider a weight function $w_{ij}$ such that it gives more weight to the points nearest to the pivot point and less weight to points far from the pivot point. At this stage we take into account a kernel function of the form $w_{ij}= F[(z_j-z_{i,0})/h)]$. The choice of weight function can be performed by guessing a behaviour where the points near to each other may be more correlated in comparison to the points which are far away. In this line of thought, we need to consider a higher weight to observations that are near to the pivot point $z_{i,0}$. For this reason we choose a tricube weight function of the form:
\begin{equation}\label{weightfn}
    w_{ij} =\left\{ \begin{array}{rcl}
    (1-u_{ij}^3)^3 & \mbox{for}
    & |u_{ij}|<1 \\ 0 & \mbox{for} & \text{  otherwise}\,, \\
    \end{array}\right.
\end{equation}
where $u_{ij}$ is defined by
\begin{equation}\label{uij}
    u_{ij} \equiv \frac{z_{i,0}-z_j}{h}\,.
\end{equation}
Here $h= \text{max}|z_{i,0}-z_j|$ is the maximum distance between the point of interest and the $j^{\text{th}}$ element of its window. 

\item We perform a fit of the subset of the data to a local polynomial $f(\theta_i)$, with $\theta_i$ being the free parameters of the polynomial model, up to first or second order using the weight function $w_{ij}$. The main idea under the LOESS assumption is to fit the small subset of data using low degree polynomials. In another case, a higher degree  polynomial may increase the computational cost without giving any significant improvement on the result. Therefore, the polynomial fit for each subset of the data is usually of first or second degree i.e. linear or quadratic. In this work we consider a fit neighbourhood subset of each pivot point with a linear polynomial fit as:
$f_i(a,b)= a+bz_i$, which allows to fit an
extrapolant function to the averaged and error-contaminated estimates decrease.
Also, we define a $\chi^2$-statistic over this subset of data as
\begin{equation}\label{eq:chi2-loess}
    \chi^2_i=\sum_{j=1}^{n} w_{ij} (\eta^{\text{obs}}_{j}-f_j(\theta_i))^2\,.
\end{equation}

\item As it is standard, we minimise Eq.~(\ref{eq:chi2-loess}) to obtain the best fit value(s) $\theta_{i}$. Then, the reconstructed value of $\eta^{obs}_i$ at pivot point $z_{i,0}$ is given by
\begin{equation}\label{reconstreta}
    \widehat{\eta}_i= f_i(\theta_0)\,,
\end{equation}
\end{enumerate}
where $\theta_0$, denotes best value at the pivot point.
We repeat these four steps for each point in the data set until we obtain the corresponding reconstructed value at each selected point.  Furthermore, in LOESS we require to fix three numerical parameters: (i) the bandwidth or smoothing parameter, (ii) the weight function and (iii) the degree of the local polynomial. To select viable values of each one we need to consider the following:

\noindent Once the LOESS process is done, we proceed with the Simex method. In cosmology most of the observed quantities come with some degree of noise or measurement errors. However, in the LOESS method we do not  use the observed measurement  errors $\sigma_i$ while reconstructing the response parameter.
The effect of observational errors  can be accommodated in the LOESS process by using a statistical technique called Simex. This method is based on a two step resampling approach. In the simulation step,  some  additional error is introduced by hand in the data with some  controlling parameter $\xi$. Next, by using regression analysis on this new  data set, we try to trace the effect of  the measured error in our original. To compute this we consider three steps:
\begin{enumerate}[(a)]
\item For  the simulation step, we  introduce a fixed amount of measurement error in each observation data point  and define  a new variable as
\begin{equation}
\label{eq:simex-eta}
    \eta^{\text{rec}}_i(\eta_k)= \eta^{\text{obs}}_i + \sqrt{\eta_k} \,\,\,\sigma_i\,,
\end{equation}
where $\sigma_i$ is the measurement error associated with the observed data $\eta^{\text{obs}}_i$ (we assume that the data points are independent of each other). Here the  parameter $\eta_k$ acts as the controlling parameter for the variance of the measurement error. It is a vector of length $K$ such that $\eta_k>0$. We thus   form a matrix of order $K \times N$. The elements of this matrix will be the values of $\eta^{\text{rec}}_i(\eta_k)$ where $i= 1,2 \ldots N $ and $k= 1,2 \ldots K $. Notice that we assume a standard normal distribution of the errors.

\item After, we reconstruct each data point given in each row of the matrix by applying the LOESS technique and 
by selecting each column of the reconstructed values of $\eta^{\text{rec}}$,  we can apply the simple regression technique to find the best fit value. At this stage, a quadratic polynomial is a better choice for our purposes.

Finally,  we obtain a row having $N$ elements. Each $\widehat{\eta}^{\text{rec}}_i(\eta_k)$ can be written as a function of $\eta_k$, i.e.
\begin{equation}\label{eq:simex-eta2}
    \widehat{\eta}^{\text{rec}}_i(\eta_k)= k_1+k_2 \eta_k+k_3 \eta_k^2\,,
\end{equation}
where $k_1$, $k_2$ and $k_3$ are normalised constants.
If a normal distribution of the errors is assumed, then the error variance associated with the simulated data points $\widehat{\eta}^{\text{rec}}_i$ will be $(1+\eta_k)\sigma_i^2$. On substituting  $\eta_k=-1$, we obtain error-free smoothed data points.

\item To construct the 1-$\sigma$ and 2-$\sigma$ C.L. around the nonparametric regression curve of $\widehat\eta$ , we assume that the errors are distributed normally. This can be obtained by using the limiting values: $\widehat \eta_i\pm \gamma\sqrt{V(\widehat \eta_i)}$, with $\gamma=1,2,\ldots$ and
\begin{equation} \label{eq:var}
    V(\widehat{\eta}_i) = \frac{1}{df_{\text{res}}}\sum_{i}^{N} d_i^2\sum_{j}^{N}w_{ij}^2\,,
\end{equation}
where $d_i = \eta_i - \widehat\eta_i$ and  $df_{\text{res}}= N- df_{\text{mod}}$, such that $df_{\text{mod}}$ is the number of effective degrees of freedom or the effective number of parameters used in this regression. We calculate it by using the normalised smoothing matrix \textbf{S}, such that  $df_{\text{mod}}= \text{Tr}($\textbf{SS}$^T$). The smoothing matrix \textbf{S}, which is a $ N \times N$ square matrix of $w_{ij}$ elements,  is  directly calculated from the weight function.
\end{enumerate}

\noindent LOESS-Simex reconstruction has been used to reconstruct the cosmic expansion in Ref.~\cite{Montiel:2014fpa}
along with the transition redshift reconstruction of the cosmic accelerated expansion cosmokinematics analysed in Ref.~\cite{Rani:2015lia}. Furthermore, its implementation to null tests of dark energy has been done in order to perform an
Om diagnostic where a static domain wall network dark energy scenario results viable \cite{Escamilla-Rivera:2015odt}. Also, in Ref.~\cite{Fernandez-Hernandez:2018yao} it was shown that LOESS-Simex is practical in astrophysical configurations to reconstruct rotation curves for dark matter profiles .


\subsection{Gaussian Processes reconstruction for cosmology}\label{sec:GP_intro}

GP regression is a non-parametric approach to reconstructing observational data which are incorporated as a finite collection of normally distributed points. This type of supervised learning is implemented for stochastic data whose data points are related together enough such that a covariance function can be fit in such a way as to prescribe these relations in a specific way \cite{10.5555/971143,10.5555/1162254}. In practice a GP is defined in terms of the mean function $\mu(z)$ together with its associated two--point covariant function $\mathcal{C}(z,\tilde{z})$ which produces the continuous realization
\begin{equation}
	\xi(z)\sim\mathcal{GP}\left(\mu(z),\mathcal{C}(z,\tilde{z})\right)\,,
\end{equation}
where the uncertainty $\Delta\xi(z)$ is also part of the reconstruction which produces a realization region $\xi(z) \pm \Delta \xi(z)$. By and large the mean is set to zero without loss of generality so that the covariance function becomes the defining feature for each reconstructed data set. For redshift points $z^*$ which represent points where observations do not occur, we can define a kernel function for the covariance function such that $\mathcal{K}\left(z^*,z^{*'}\right) = \mathcal{C}\left(z^*,z^{*'}\right)$ where the trained covariance function is utilised. Thus, the kernel will embody all the information about the strength and amplitude of the correlations between the redshift data points \cite{Gomez-Valent:2018hwc}. The only strong requirement on the kernel is that it is a symmetric function.

For observational point $\tilde{z}$, we also have available to us information on the associated uncertainty region covariance matrix $\mathcal{D}\left(\tilde{z},\tilde{z}'\right)$ between data points. Thus, the covariance function for these points can be written as $\mathcal{C}\left(\tilde{z},\tilde{z}'\right) = \mathcal{K}\left(\tilde{z},\tilde{z}'\right) + \mathcal{D}\left(\tilde{z},\tilde{z}'\right)$ which will provide information to fit the kernel. Naturally, observational and reconstructed points can be interrelated by the kernel alone through $\mathcal{C}\left(z^*,\tilde{z}'\right) = \mathcal{K}\left(z^*,\tilde{z}'\right)$ \cite{Busti:2014aoa}.

Therefore, for a Gaussian distribution, the posterior distribution of a reconstructed function can be expressed via the joint Gaussian distribution of different data points in which the kernel expresses the interrelated mean value and uncertainties of each point in the resulting distribution \cite{Seikel:2013fda,Gomez-Valent:2018hwc}. In Appendix \ref{appendix:GP} we develop a discussion of this GP technique and the GaPP code\footnote{\url{http://ascl.net/1303.027}} in order to check how sensitive the chosen horizontal axis scale is related to thedata points.
In the present work, we use the squared--exponential kernel but there exist a plethora of such kernel options \cite{10.5555/1162254}. Despite each producing a slightly different non-parametric reconstruction of a distribution, they largely agree to well within $1\sigma$ confidence regions. The general purpose squared--exponential kernel is explicitly defined as
\begin{equation}\label{eq:square_exp}
	\mathcal{K}\left(z,\tilde{z}\right) = \sigma_f^2 \exp\left[-\frac{\left(z-\tilde{z}\right)^2}{2l_f^2}\right]\,,
\end{equation}
where $\sigma_f$ and $l_f$ are the kernel hyperparameters, which are parameters of the kernel but do not parameterise the function being reconstructed, as a model would. The hyperparameters characterise the variance in the data through $\sigma_f$ and the length--scale through $l_f$. Thus, the hyperparameters define the smoothness and reach for fluctuations in a signal. As a result, larger values of $l_f$ lead to smoother GP functions while higher values of $\sigma_f$ express a lower signal--to--noise ratio. 

To determine the best suited hyperparameters, their values are derived from the maximisation of the probability of the GP to generate the data set under consideration which is implemented via the minimisation of the GP marginal likelihood similar to a Bayesian approach \cite{Busti:2014dua,Verde:2014qea,Seikel:2013fda,Li:2015nta}. GP regression has now been extensively studied as a tool for reconstructions in the cosmological context which have largely focused on the late-time behaviour of the expansion rate \cite{Shafieloo:2012ht,Seikel:2013fda,Cai:2015zoa,Cai:2015pia,Wang:2017jdm,Zhou:2019gda,Mukherjee:2020vkx,Gomez-Valent:2018hwc,Zhang:2018gjb,Aljaf:2020eqh,Li:2019nux,Liao:2019qoc,Yu:2017iju,Yennapureddy:2017vvb}. In Refs.~\cite{Briffa:2020qli,Cai:2019bdh} the use of GP reconstructions was extended from kinetic parameters to theories beyond $\Lambda$CDM.
As an example in this line of thought, $f(T)$ gravity was then extended into the perturbative section in Ref.~\cite{LeviSaid:2021yat}, where an $f\sigma_8$ data set was used to constrain the form of potential $f(T)$ cosmology models that are in agreement with earlier GP reconstructions of growth data in Ref.~\cite{Benisty:2020kdt}. Other recent work also includes a possible resolution to the Hubble tension problem within $\Lambda$CDM in Refs.~\cite{Krishnan:2020vaf,Colgain:2021ngq} where the model independent nature of GP is brought into question due to its dependence on the kernel choice in the regression procedure of the GP reconstruction algorithm. Furthermore, in quintessence and k-essence models from Horndeski theories of gravity \cite{Reyes:2021owe} that represent an extension to quintessence can reproduce reconstructions of the late expansion of the Universe within 2$\sigma$.


\section{Reconstructions applied to cosmic data}
\label{sec:results}

With the recipes explained above, in this section we develop the reconstruction for late-time observations based on luminosity distances and Hubble flow reconstructions.
The LOESS-Simex and GP reconstructions are based on current data sets of Type Ia supernovae (SNeIa), observational Hubble data (CC) and baryon acoustic oscillations (BAO). 


\subsection{Cosmic chronometers and baryonic acoustic oscillations}\label{sec:CC_BAO_data}

For the observational Hubble data we consider a sample of $l=51$ measurements in the range $0.07< z < 2.36$. A total sample size of 31 data points from passive galaxies, or cosmic chronometers (CC) \cite{Moresco:2015cya,Moresco:2016mzx,2010JCAP...02..008S,2012JCAP...08..006M,2014RAA....14.1221Z}, and 20 data points which are estimated from BAO data points \cite{BOSS:2013igd,BOSS:2014hwf,Bautista:2017zgn,BOSS:2013rlg,2012MNRAS.425..405B,2017MNRAS.470.2617A,10.1093/mnras/stt1290,BOSS:2016zkm,Oka:2013cba,Gaztanaga:2008xz} under a $\Lambda$CDM prior. Following Ref.~\cite{Magana:2017nfs}, we assume that these points are not correlated. To remove the model dependence in some of the data points, we use a sound horizon $r_s$ computed from Planck 2018 \cite{Aghanim:2018eyx} (henceforth denoted by PL18). To perform the model independent reconstruction, we set the construction of a $\chi_H^2$ as 
\begin{equation}
    \chi_H^2=\sum_{i=1}^{l}\frac{\left[H_{\text{recons}}\left(z_i,\mathbf{x}\right)-H_{\text{obs}}(z_i)\right]^2}{\sigma^2_H(z_i)}\,,
\end{equation}
where $\text{H}_{\textit{obs}}(z_i)$ is the observed value at $z_i$, $\sigma_H(z_i)$ are the observational errors, and $\text{H}\left(z_i,\mathbf{x}\right)$ is the reconstructed H for the same $z_i$ with the specific parameter vector $\mathbf{x}$. 

For the BAO sample we consider 15 transversal measurements in a range of $z=[0.11,\,2.225]$
, obtained in a quasi--model independent approach. As it is standard, this can be computed by considering the 2-point angular correlation function tracers with $D_A (z; r_{\text{drag}})$ \citep{Nunes:2020hzy}. This data can be compared with the sound horizon today to the sound horizon at the time of recombination inferred from the CMB anisotropy measurements. Defining the fundamental quantities for this observation, we have the BAO distances $d_{z} \equiv \frac{r_{s}(z_{d})}{D_{V}(z)}$, with  $r_{s}(z_{d}) = \frac{c}{H_{0}} \int_{z_{d}}^{\infty} \frac{c_{s}(z)}{E(z)}\, \mathrm{d}z$ being the comoving sound horizon at the baryon dragging epoch, $c$ the speed of light, $z_{d}$ is the drag epoch redshift, and $c^{2}_{s}= c^2/3[1+(3\Omega_{b0}/4\Omega_{\gamma 0})(1+z)^{-1}]$ the sound speed with $\Omega_{b0}$ and $\Omega_{\gamma 0}$ being the current values of baryon and photon density parameters, respectively. We can also define the dilation scale by
\begin{equation}
    D_{V}(z,\Omega_{m0}; \Theta) = \left[   \frac{c \, z (1+z)^2 D_{A}^2}{H(z, \Omega_{m0}; \Theta)} \right]^{1/3}\,,
\end{equation}
where $D_{A}$ is the angular diameter distance given by
\begin{equation}
    D_{A}(z,\Omega_{m0}; \Theta) = \frac{1}{1+z} \int_{0}^{z} \frac{c \, \mathrm{d}\tilde{z}}{H(\tilde{z}, \Omega_{m0};\Theta)} \,,
\end{equation}
where $\Theta$ is the vector that contains the free cosmological parameters to be reconstructed (in such cases fitted). Using the comoving sound horizon, the distance ratio $d_{z}$ can be related to the normalised Hubble parameter $h$ ({$H \doteq 100 h$)} and the critical densities $\Omega_{m0}$ and $\Omega_{b0}$. To connect the BAO measurements with SNeIa to CMB measurements (PL18), we take into account the Alcock-Paczynski distortion parameter given by
\begin{equation}
    F(z,\Theta)=(1+z)\frac{D_{A}(z,\Theta)H(z,\Theta)}{c}\,.
\end{equation}
Here we notice that by calibrating the $D_A$ from BAO, we can define
\begin{equation}
    \chi^2_{\text{BAO}}=(\Delta\mathcal{F}_{\text{BAO}})^{T}\cdotp C_{\text{BAO}}^{-1}\cdotp\Delta\mathcal{F}_{\text{BAO}}\,,
\end{equation}
where $\Delta\mathcal{F}_{\text{BAO}}$ is the difference between the data and the resulting value for $\Theta$, and $C_{\text{BAO}}^{-1}$ is the inverse of the covariance matrix of this sample. In our analysis the off-diagonal entries of the covariance matrix are taken to be zero. We do this since the LS method cannot tackle covariance matrices. However, we did perform the following analyses for GP both with and without the covariance matrix for this data set which resulted in very similar results. Thus, for consistency, we are able to take the GP analysis without a covariance matrix.

\begin{figure}[H]
\begin{center}
  \includegraphics[width=0.325\columnwidth]{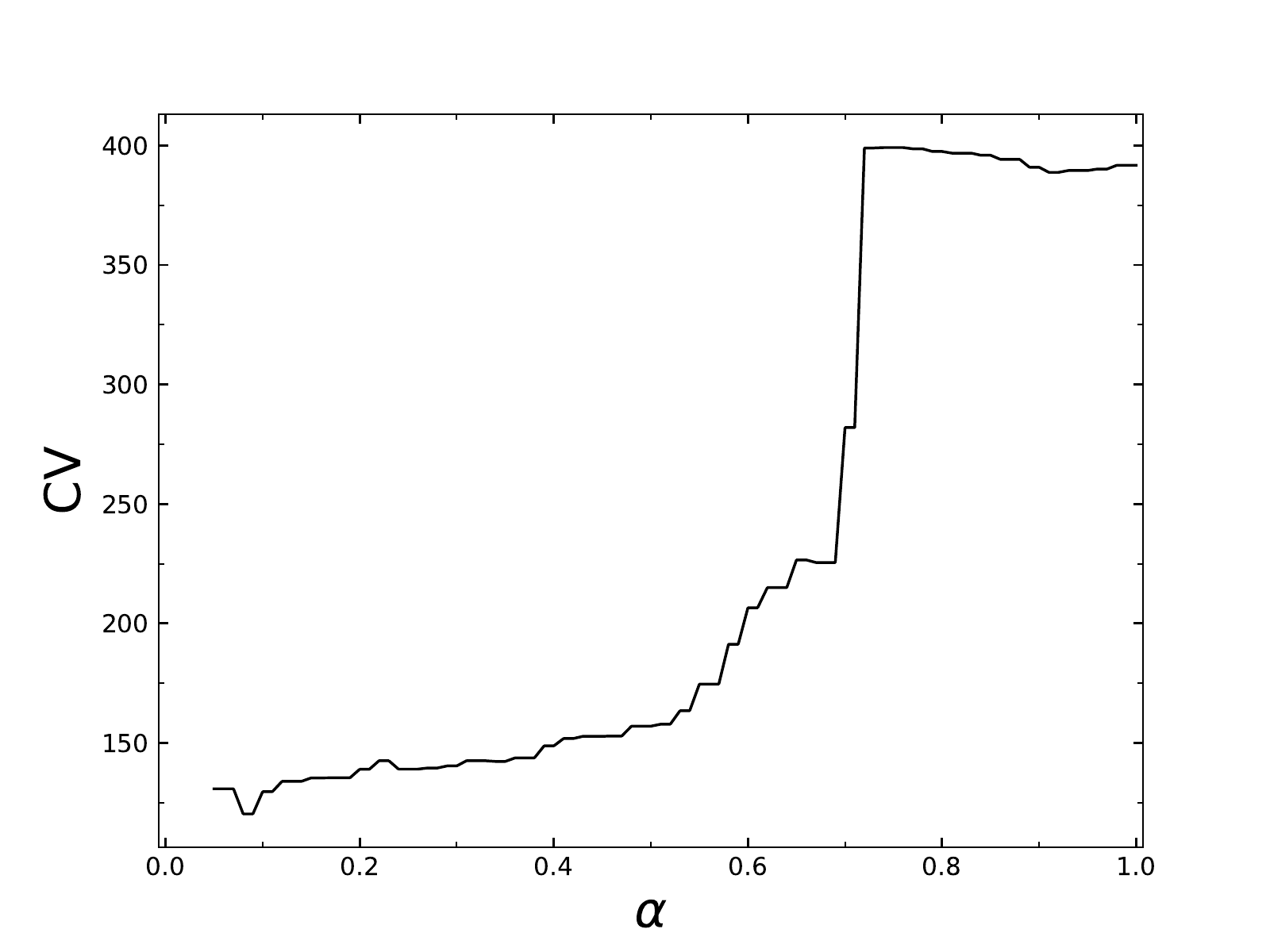}
  \includegraphics[width=0.325\columnwidth]{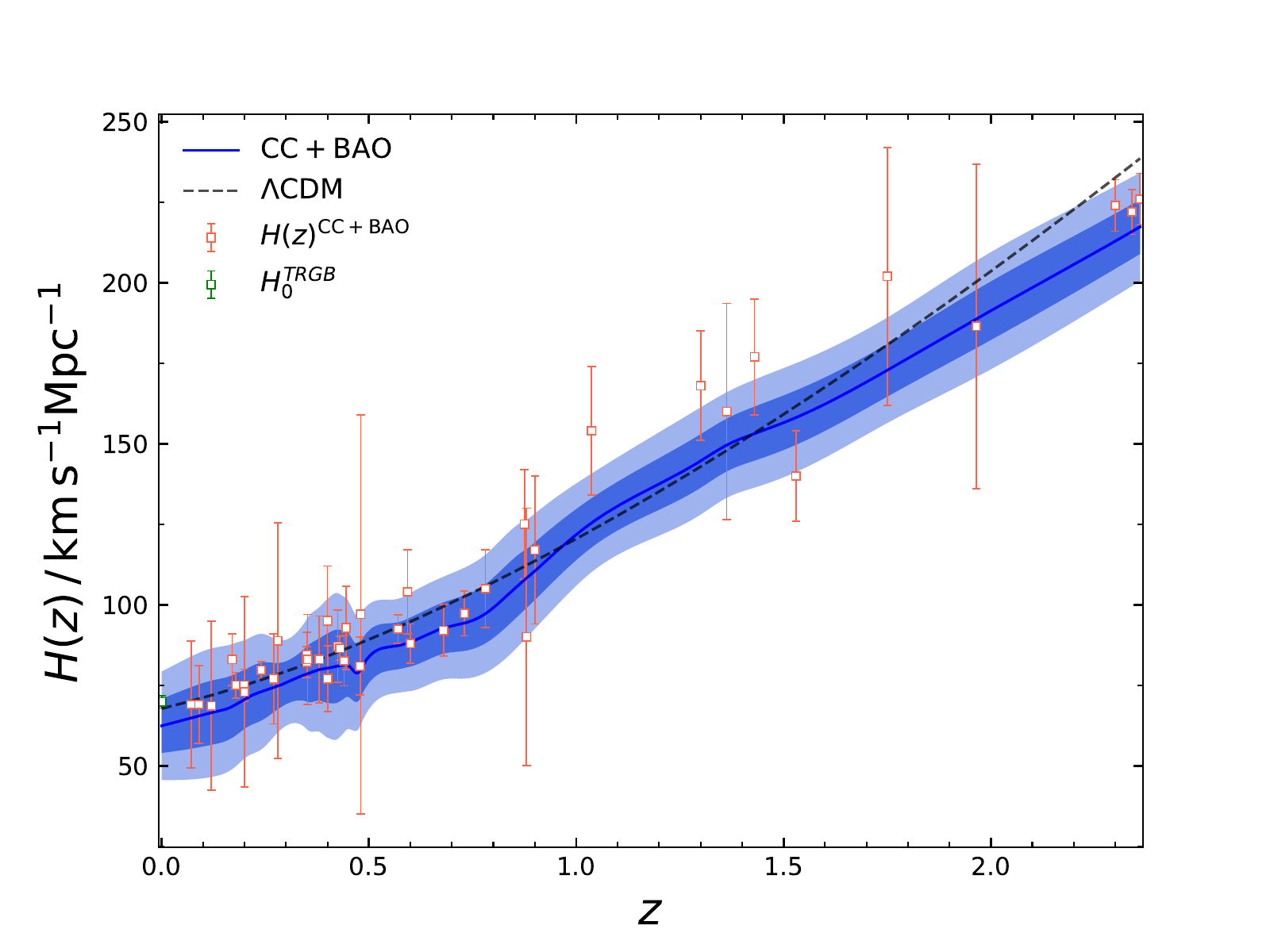}
   \includegraphics[width=0.325\columnwidth]{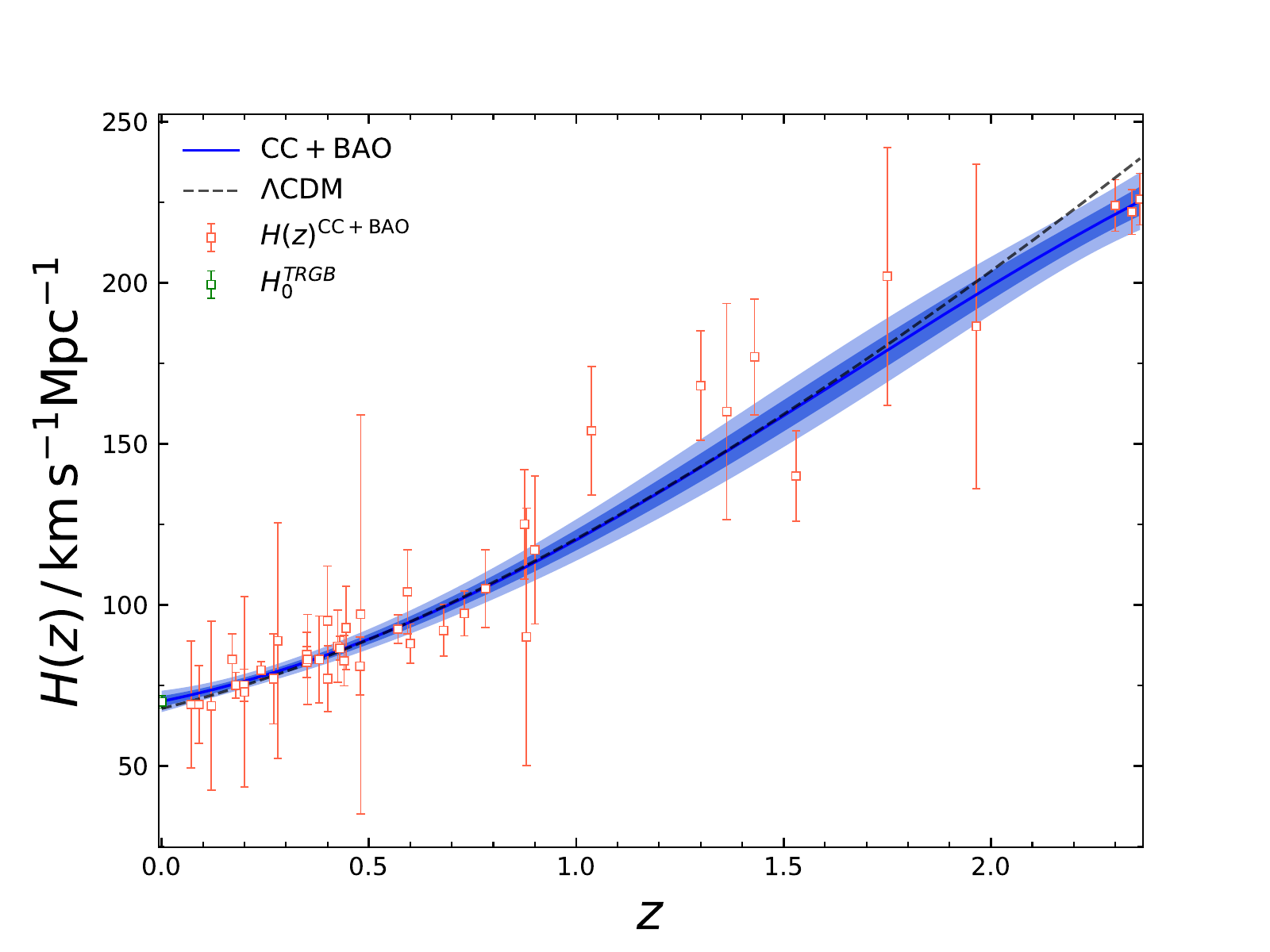}
   \includegraphics[width=0.325\columnwidth]{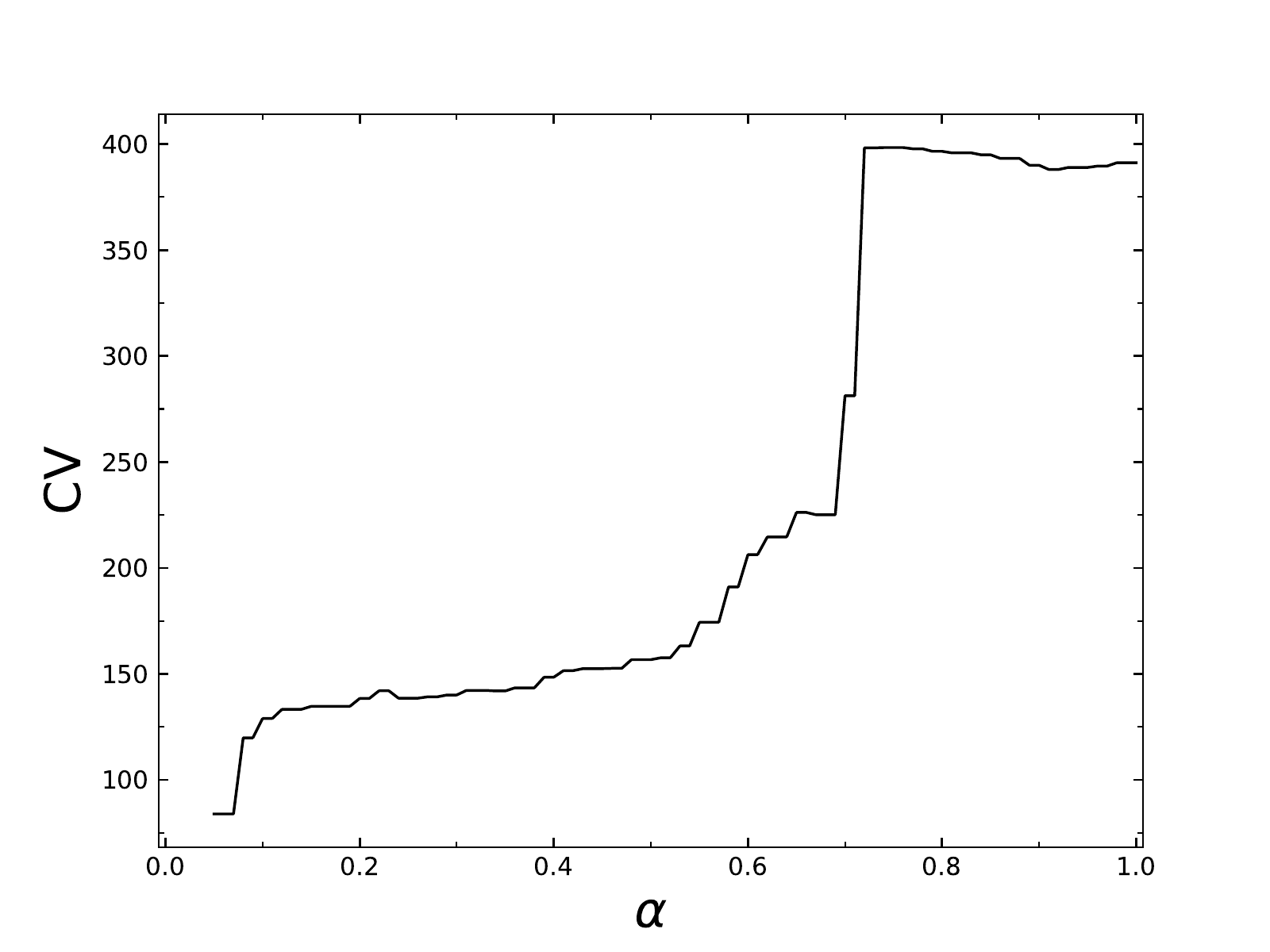}
   \includegraphics[width=0.325\columnwidth]{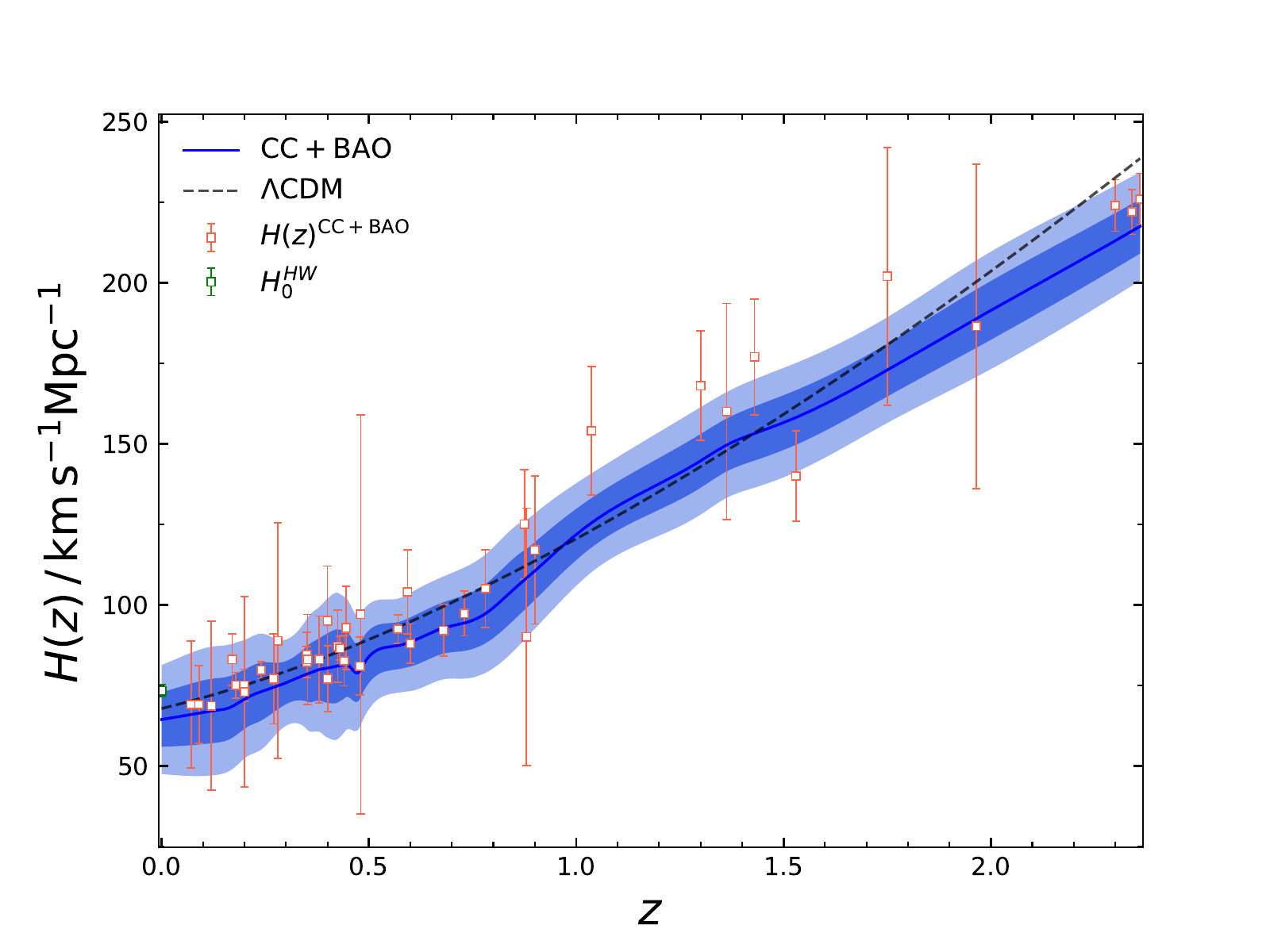}
   \includegraphics[width=0.325\columnwidth]{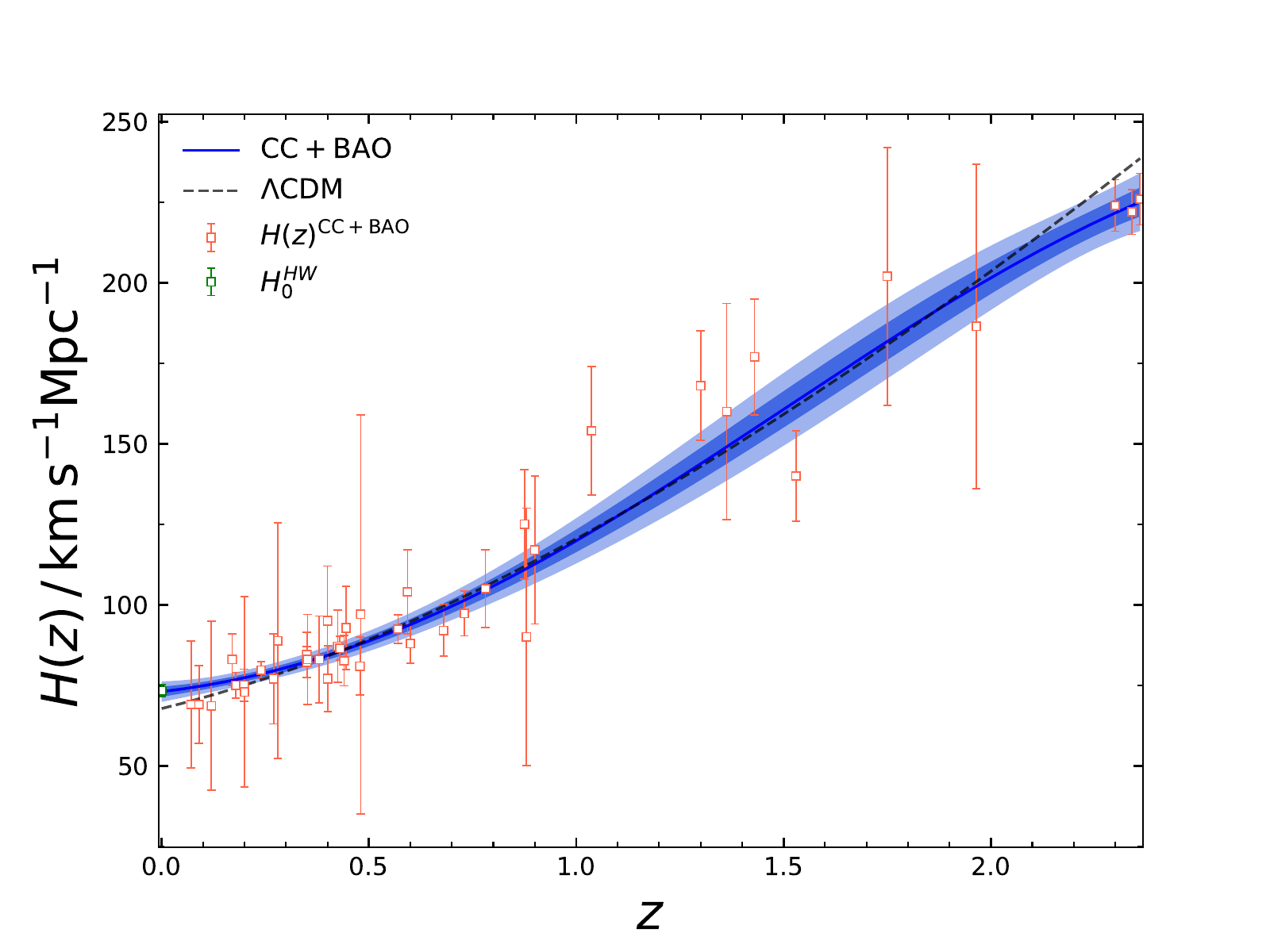}
   \includegraphics[width=0.325\columnwidth]{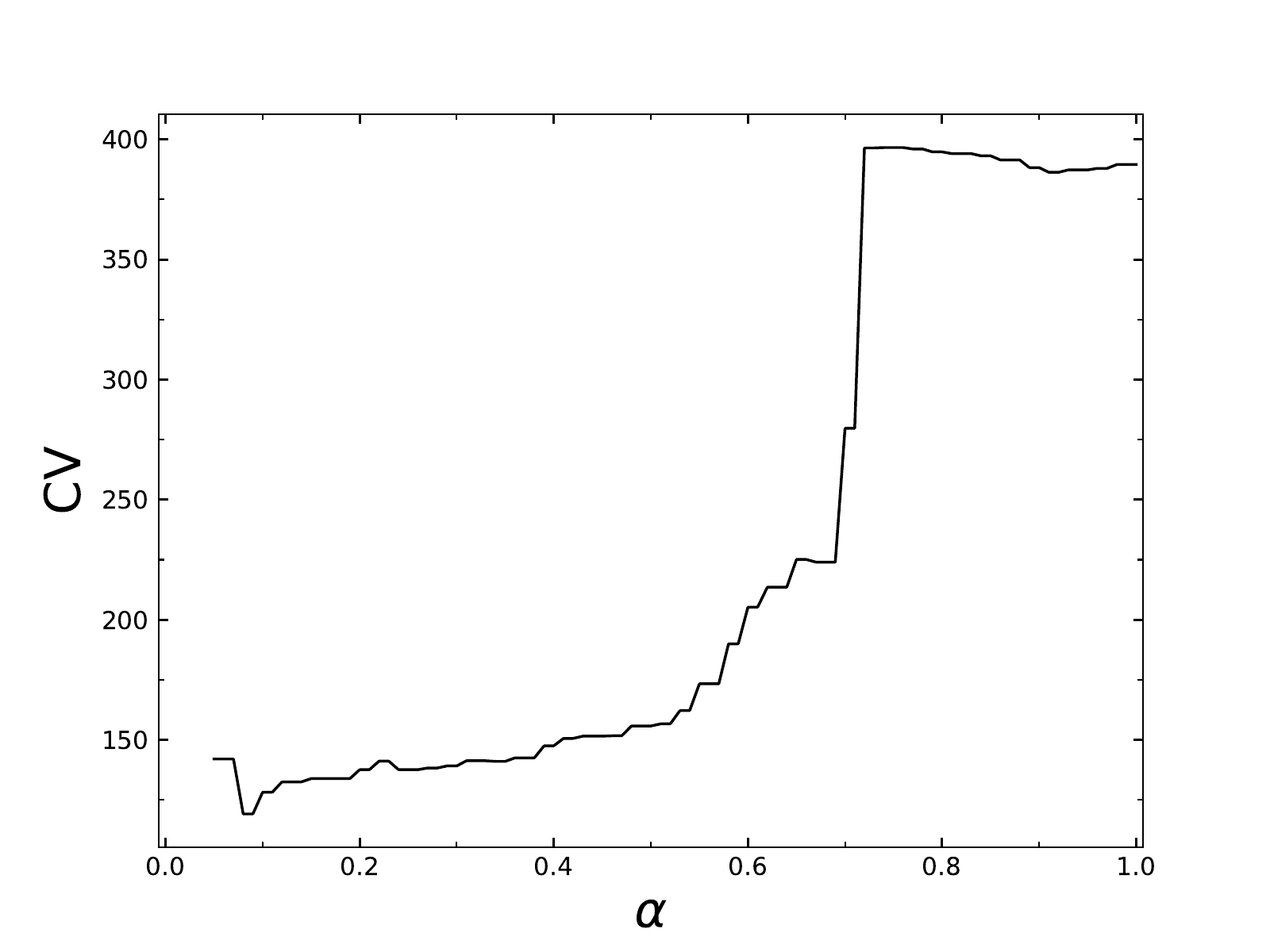}
   \includegraphics[width=0.325\columnwidth]{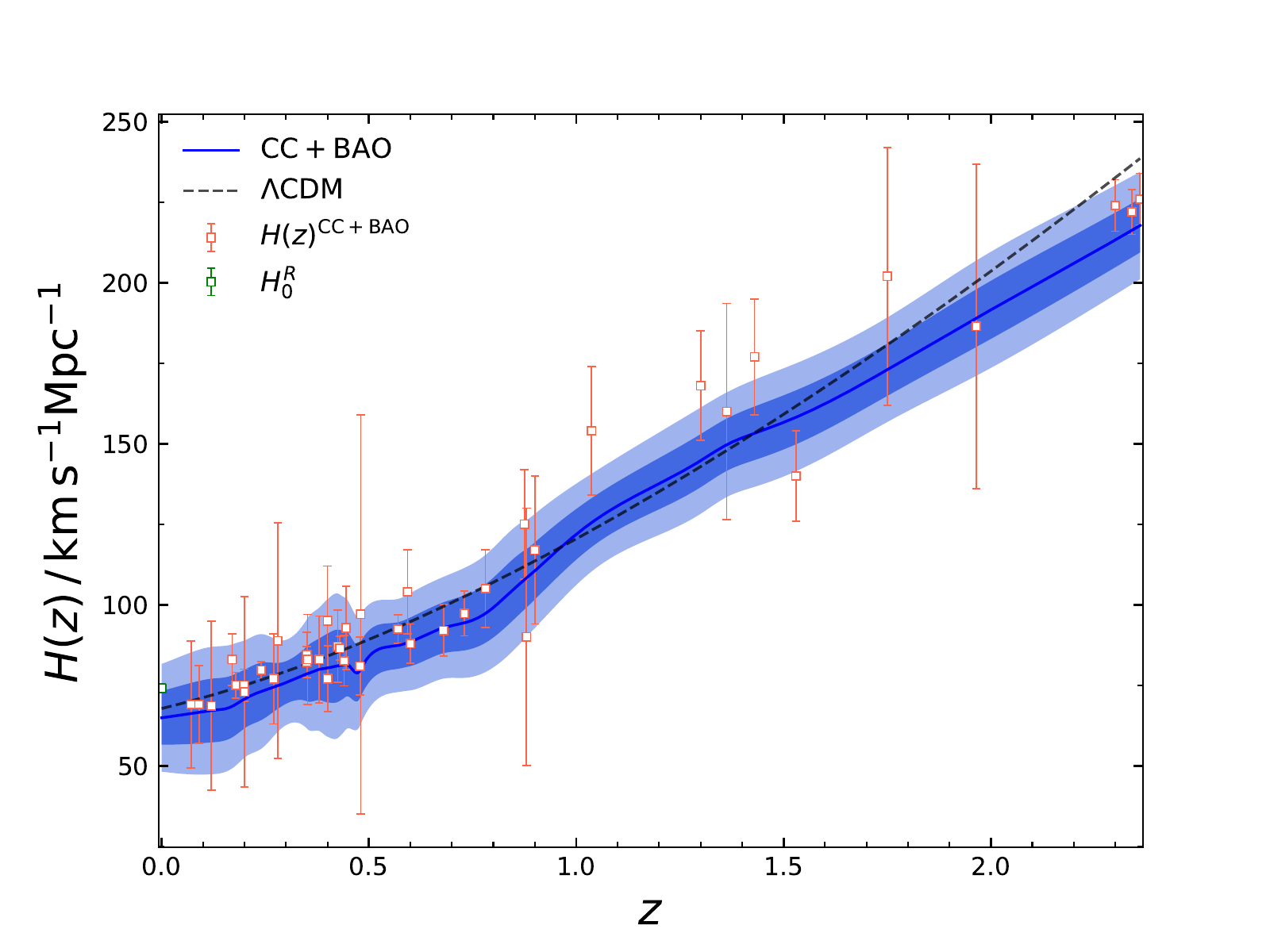}
   \includegraphics[width=0.325\columnwidth]{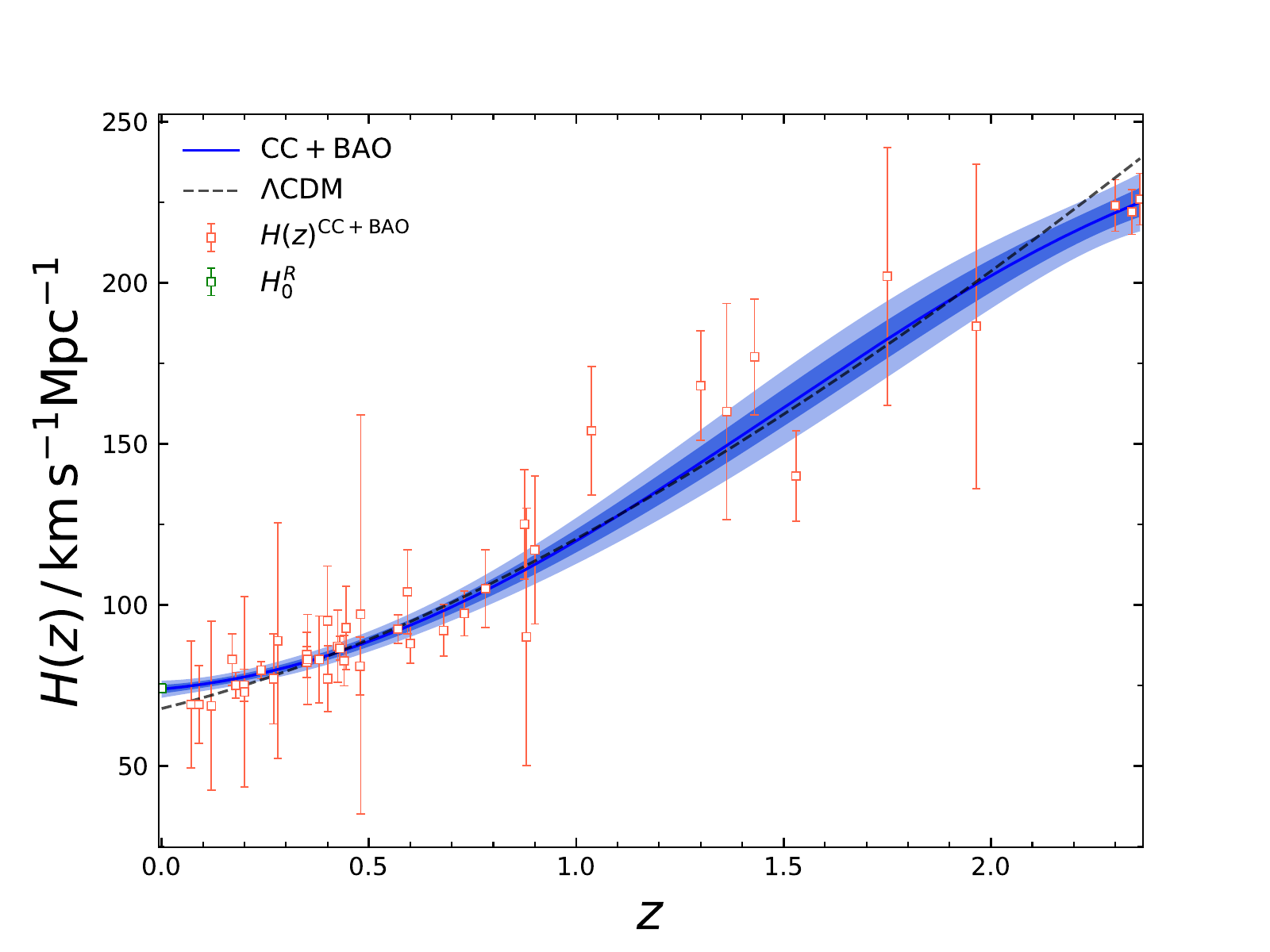}
   \caption{Analysis reconstruction using CC+BAO data set (red color points). \textit{Left column:} Cross validation (Eq. (\ref{eq:CV})) is shown for $\alpha$ values ranging between $0$ and $1$. \textit{Middle column:} The LOESS-Simex reconstructed Hubble diagram is shown (blue color contours) for each prior against a $\Lambda$CDM (PL18) reference (dashed line).  \textit{Right column:} The GP reconstructed Hubble diagram is shown (blue color contours) for each prior against a $\Lambda$CDM (PL18) reference. In both reconstructions we show the 1$\sigma$ and 2$\sigma$ confidence regions of the reconstructed data. From \textit{top} to \textit{bottom} we denote the $H_0$ prior cases as: TRGB, HW and R19, respectively.}\label{fig:CC_BAO_recon}
\end{center}
\end{figure}


\subsection{Pantheon data set}\label{sec:SN_data}

The Pantheon SNeIa compilation is one of the latest Type Ia Supernovae compilations \citep{Scolnic:2017caz} and  it contains 1048 SNeIa at redshift $0.01<z<2.26$. The constraining power of this kind of supernovae is due to the fact that this observations can be used as standarisable candles. This can be implemented through the use of the distance modulus
\begin{equation}
    \mathcal{F}(z,\Theta)_\text{theo}=5\log_{10}\left[D_L(z,\Theta)\right]+\mu_0\,,
\end{equation}
where $D_L$ is the luminosity distance given by
\begin{equation}
    D_L(\Theta)=(1+z)\int_0^z{\frac{c\, dz'}{H_0 E(z',\Theta)}}\,,
\end{equation}
and $\Theta$ is the vector with the free cosmological parameters to be fitted. We notice that the factor $c/H_0$ can be absorbed in $\mu_0$. Furthermore, we can write $\Delta\mathcal{F}(\Theta)=\mathcal{F}_{\text{theo}}-\mathcal{F}_{\text{obs}}$, using for this purpose the  distance modulus $\mathcal{F}_{\text{obs}}$ associated with the observed magnitude. At this point, it may be thought  that a possible $\chi_{SN}^2$ is given by
\begin{equation}
    \chi_{SN}^2(\Theta) = \left(\Delta\mathcal{F}(\Theta)\right)^{T}\cdotp C_{SN}^{-1}\cdotp \Delta\mathcal{F}(\Theta)\,,
\end{equation}
where $C_{SN}$ is the total covariance matrix. As in the case of the BAO data set, we tested GP both with and without the covariance matrix with very similar results being returned. For this reason, and due to the LS method not being able to handle covariance matrices, we used the GP method without the covariance matrix approach for consistency. This equation can be used to contain the nuisance parameter $\mu_0$, which in turn is a function of the Hubble constant, the speed of light $c$ and the SNeIa absolute magnitude. To circumvent this issue, $\chi_{SN}^2$ is marginalised analytically with respect to $\mu_0$ and we can obtain a new $\chi_{SN}^2$ estimator, given by (using Eq.(6,7) of Ref.~\cite{Scolnic:2017caz})
\begin{equation}
    \chi_{SN}^2(\Theta)=\left(\Delta\mathcal{F}(\Theta)\right)^{T}\cdotp C_{SN}^{-1}\cdotp \Delta\mathcal{F}(\Theta)+\ln{\frac{S}{2\pi}}-\frac{k^2(\Theta)}{S}\,,
\end{equation}
where $S$ is the sum of all entries of $C_{SN}^{-1}$. This equation gives an estimation of the precision of these data points independently of $\Theta$, and $k$ is $\Delta\mathcal{F}(\Omega_m,\Omega_r,\Omega_\Lambda)$ but weighed by a covariance matrix as follows:
\begin{equation}
    k(\Theta)={\left(\Delta\mathcal{F}(\Theta)\right)^{T}\cdotp C_{SN}^{-1}}\,.
\end{equation}
Also, for this sampler we 
are taking the nuisance parameter $M$ inside the sample. Consequently, we perform three calibrations taking into account
three different priors of $H_0$:

\begin{itemize}
 \item  TRGB \cite{freedman2019carnegie}:    $H_{0}^{\text{TRGB}} = 69.8 \pm 1.9\,\text{km/s/Mpc}$. 
    \item  H0LiCOW (HW) \cite{wong2020h0licow}: $H_ {0}^{\text{HW}} = 73.3^{+1.7}_{- 1.8}\,\text{km/s/Mpc}$. 
        \item  Riess et al (R19) \cite{Riess:2016jrr} :   $H_ {0}^{\text{R}} = 73.24 \pm 1.74 \,\text{km/s/Mpc}$. 
\end{itemize}
For this case we based our calibration on Ref.~\cite{Chiang:2017yrq} by
extrapolating the expression $M_1 = M_2 + 5 \log10(H_1/H_2)$, where the indices $1$ and $2$ denote our different values for $H_0$. Finally, our result is $\mu(z)=m-M$. A compilation of these measurements can be found in \cite{Di_Valentino_2021,DiValentino:2020zio,Freedman:2021ahq,Alestas:2021xes}.


\section{Comparative analysis of reconstruction methods}
\label{sec:comparative}

In this section, we will compare and contrast the GP and LOESS-Simex reconstruction methods in terms of their performance against each other at reproducing the observations along with their relative uncertainties. In particular, we aim to quantify the ability of both methods to reproduce the expansion data together with the uncertainties in those reconstructions of the data.

\begin{figure}[h]
\begin{center}
    \includegraphics[width=0.325\columnwidth]{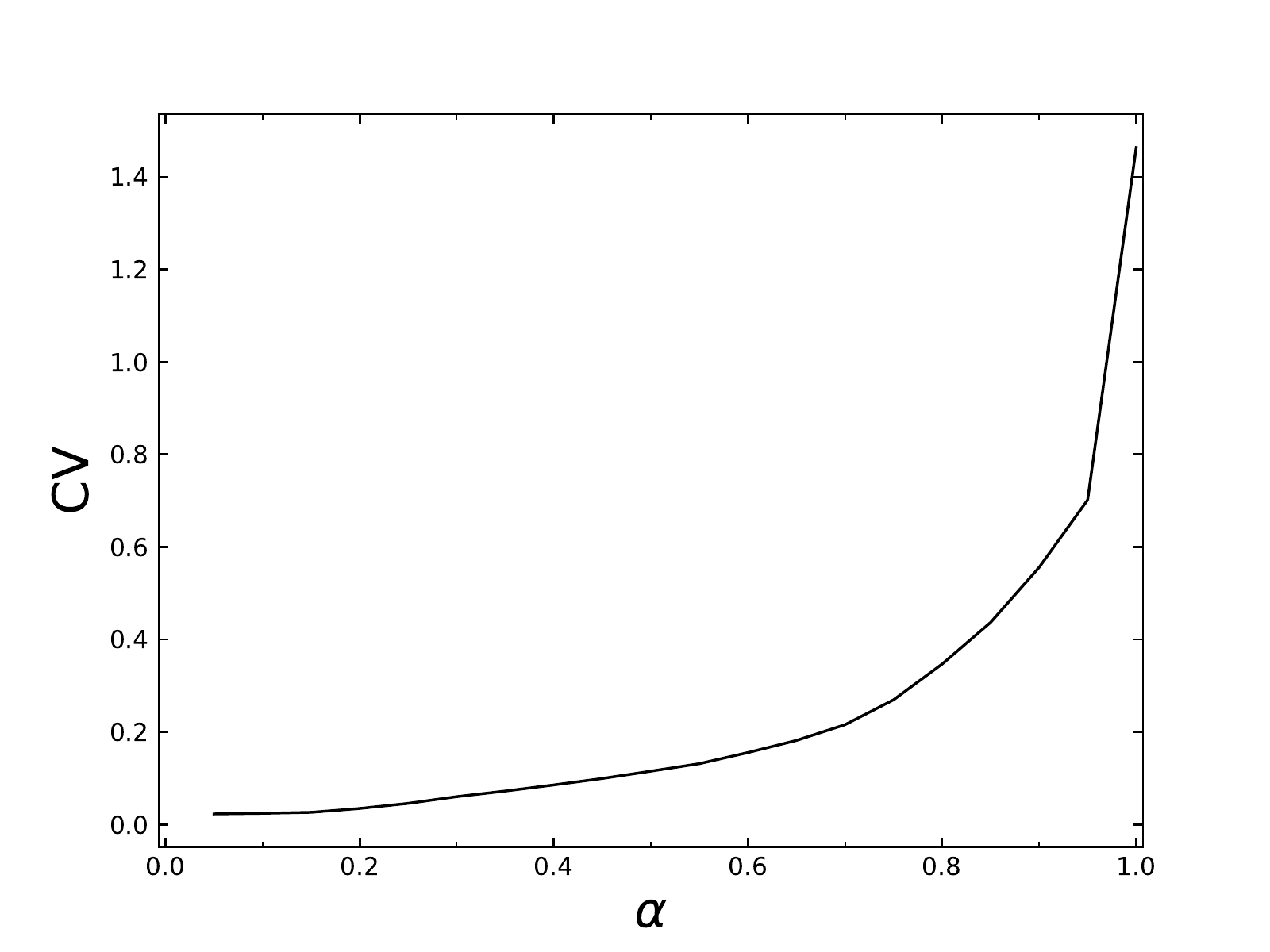}
    \includegraphics[width=0.325\columnwidth]{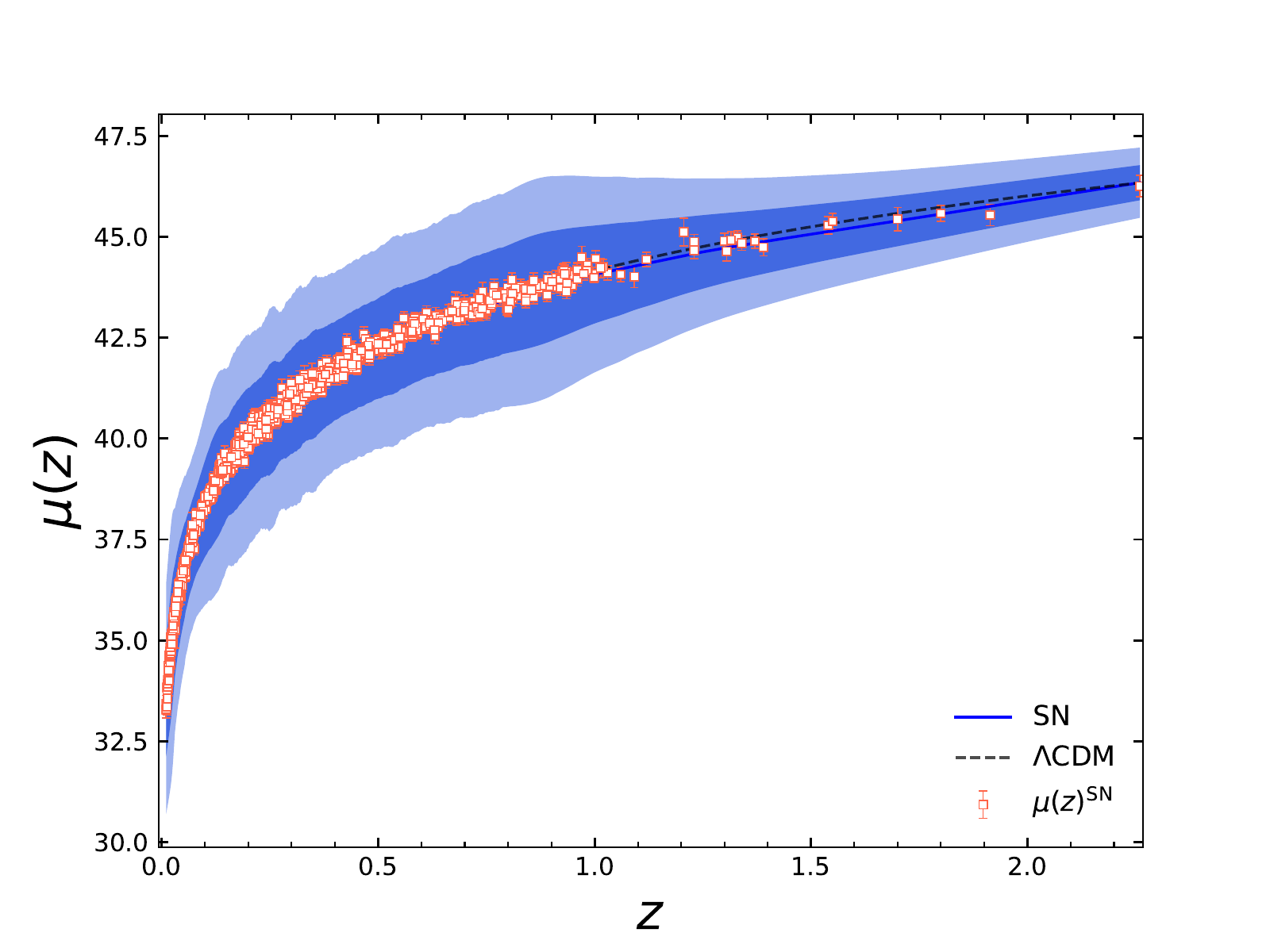}
    \includegraphics[width=0.325\columnwidth]{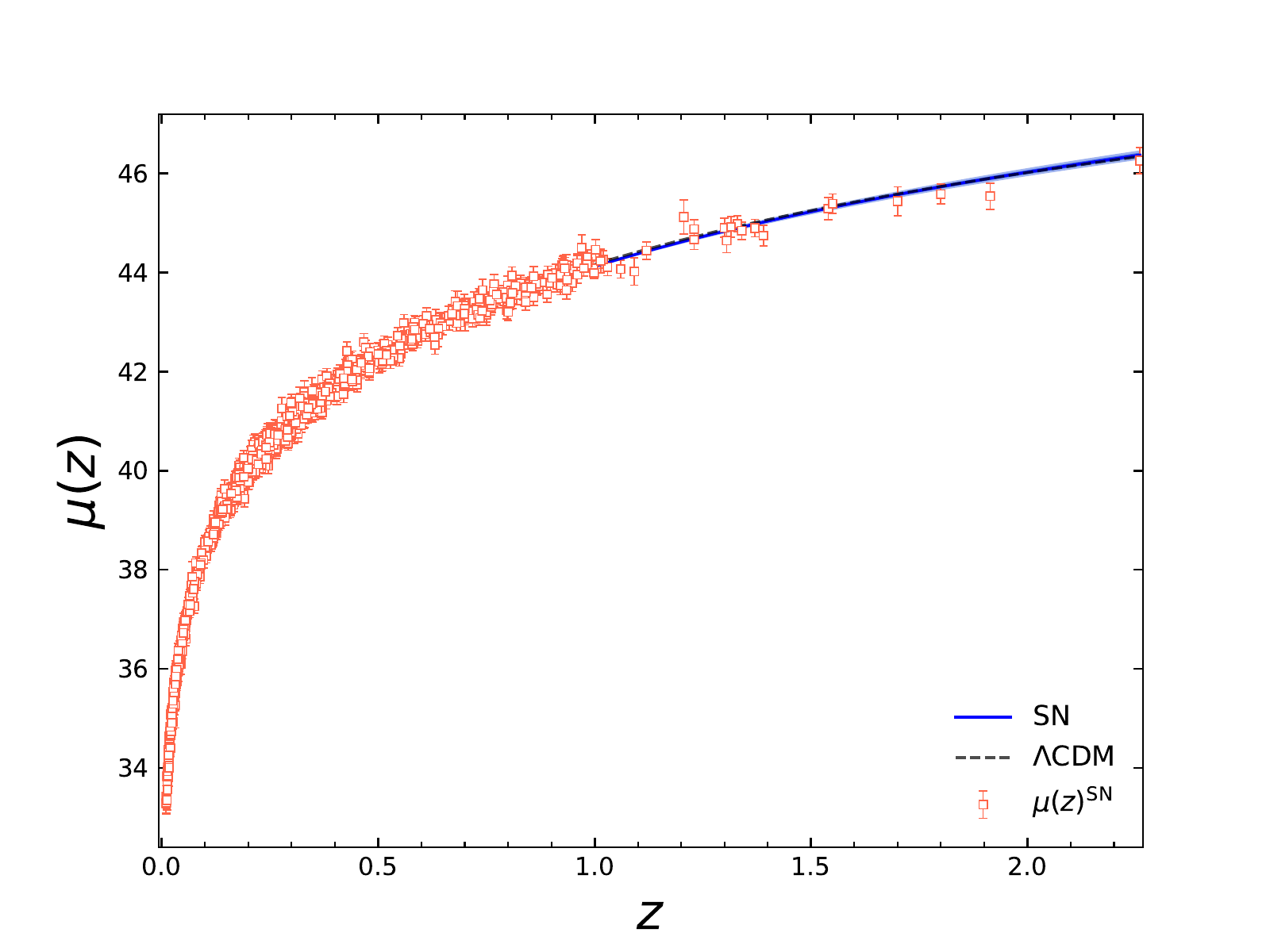}
  \includegraphics[width=0.325\columnwidth]{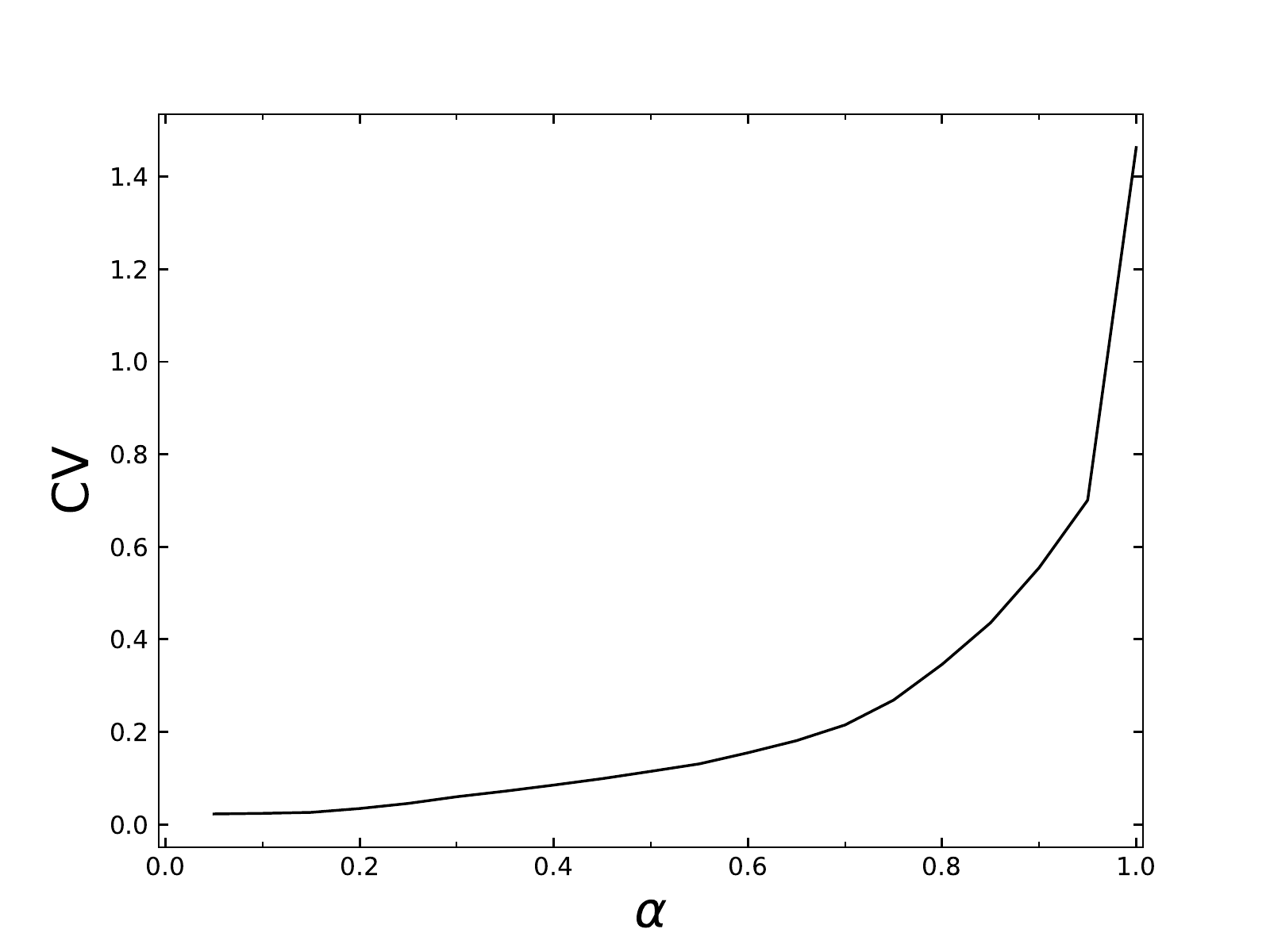}
   \includegraphics[width=0.325\columnwidth]{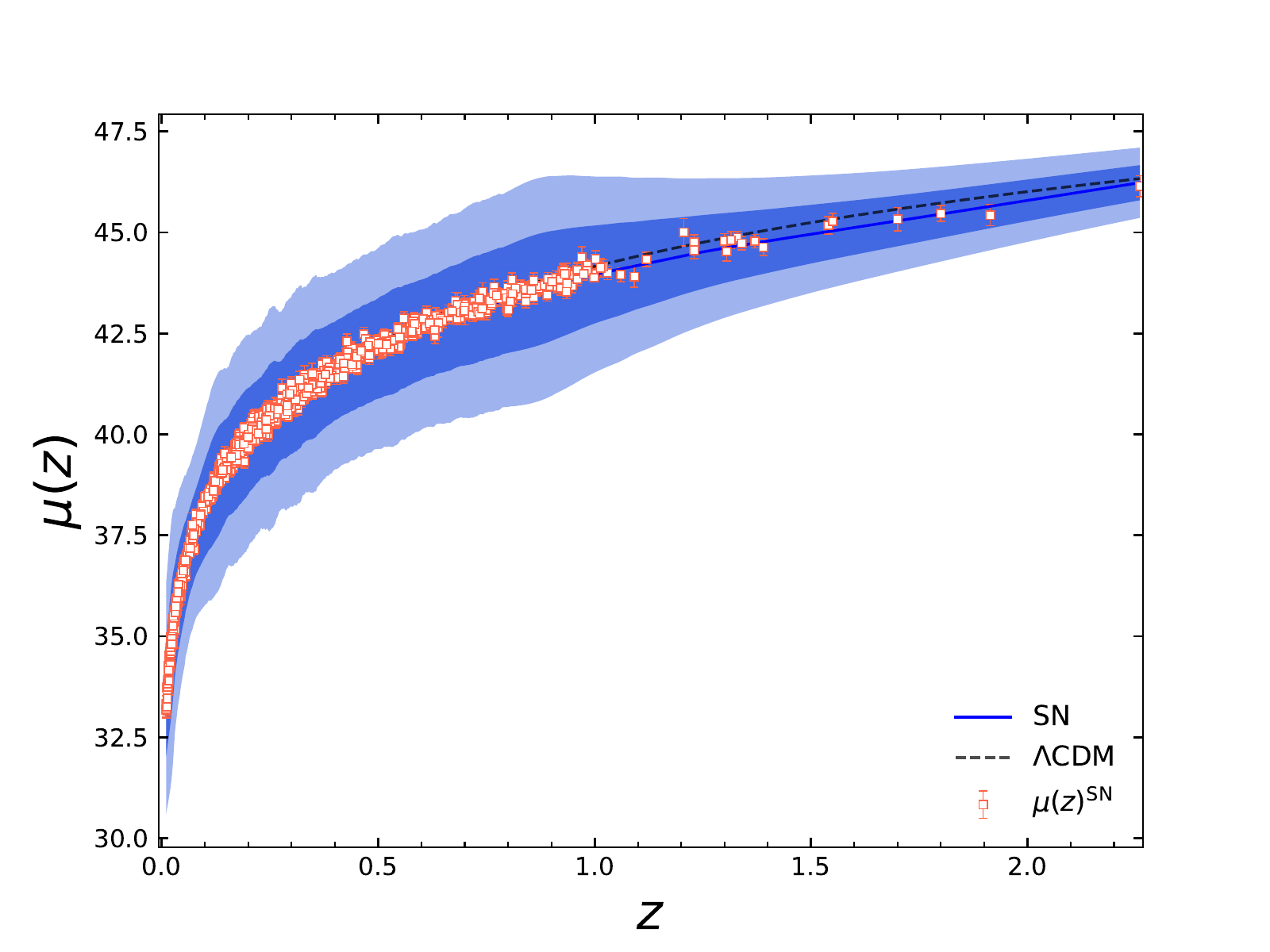}
   \includegraphics[width=0.325\columnwidth]{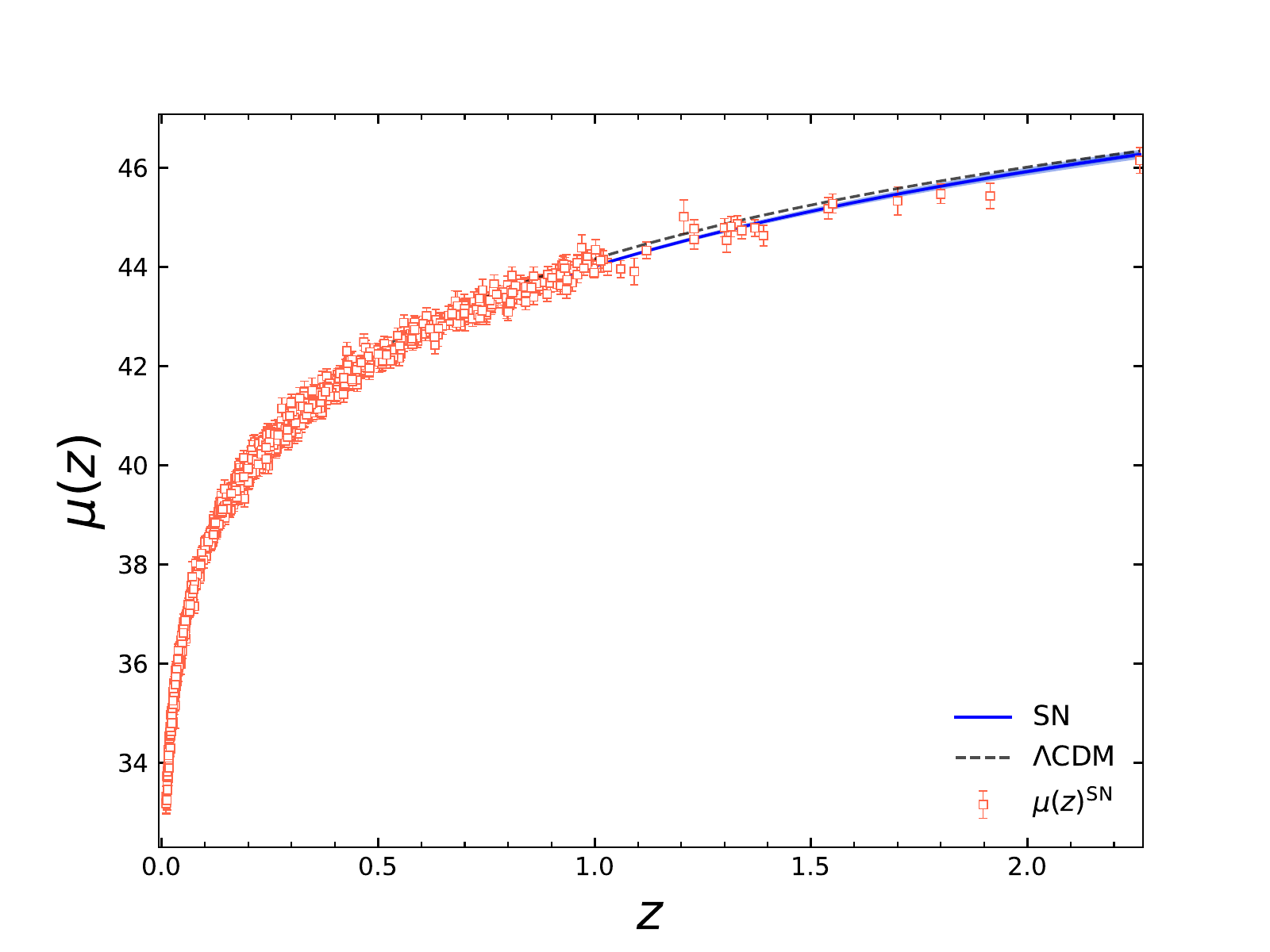}
   \includegraphics[width=0.325\columnwidth]{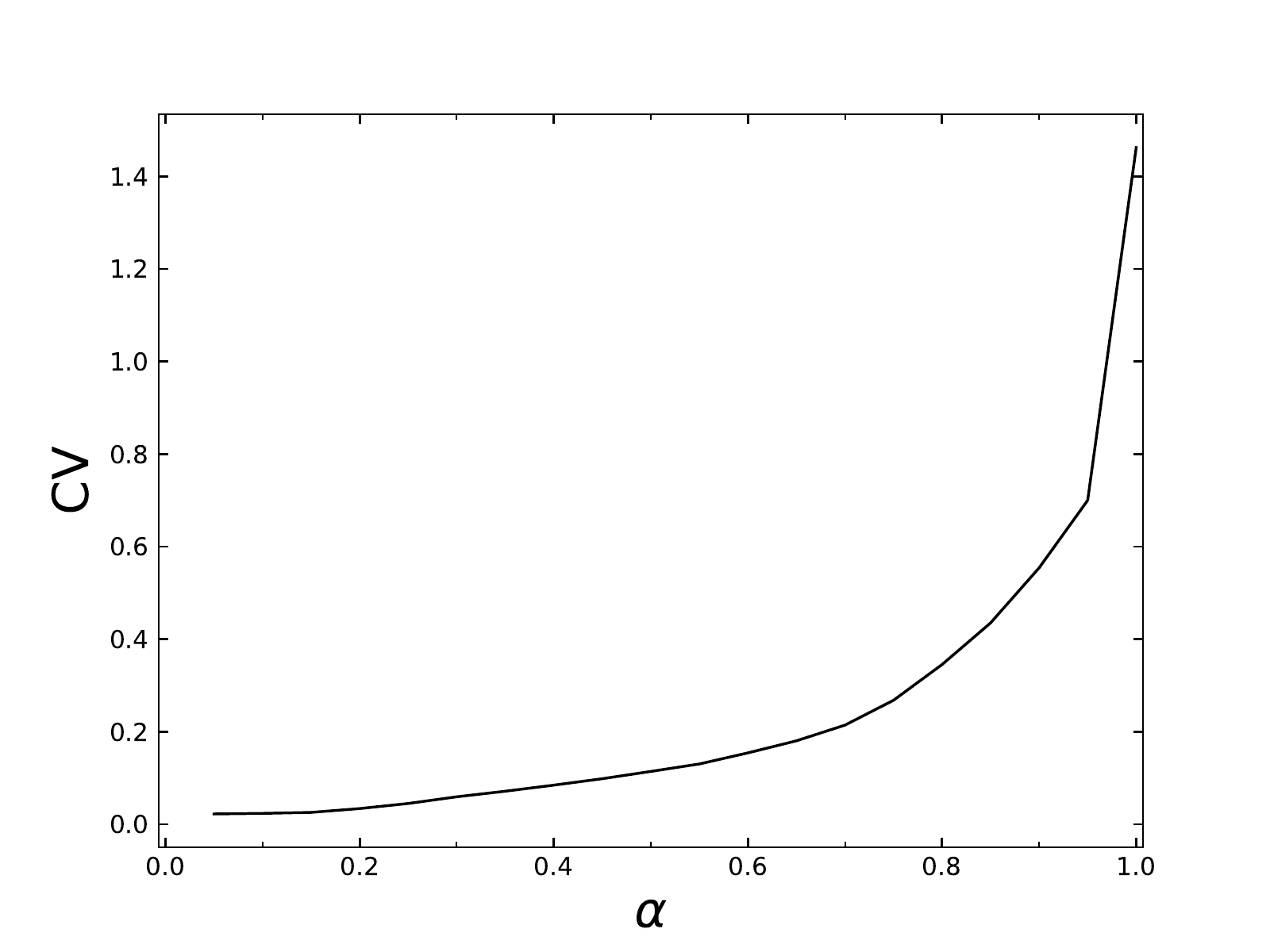}
   \includegraphics[width=0.325\columnwidth]{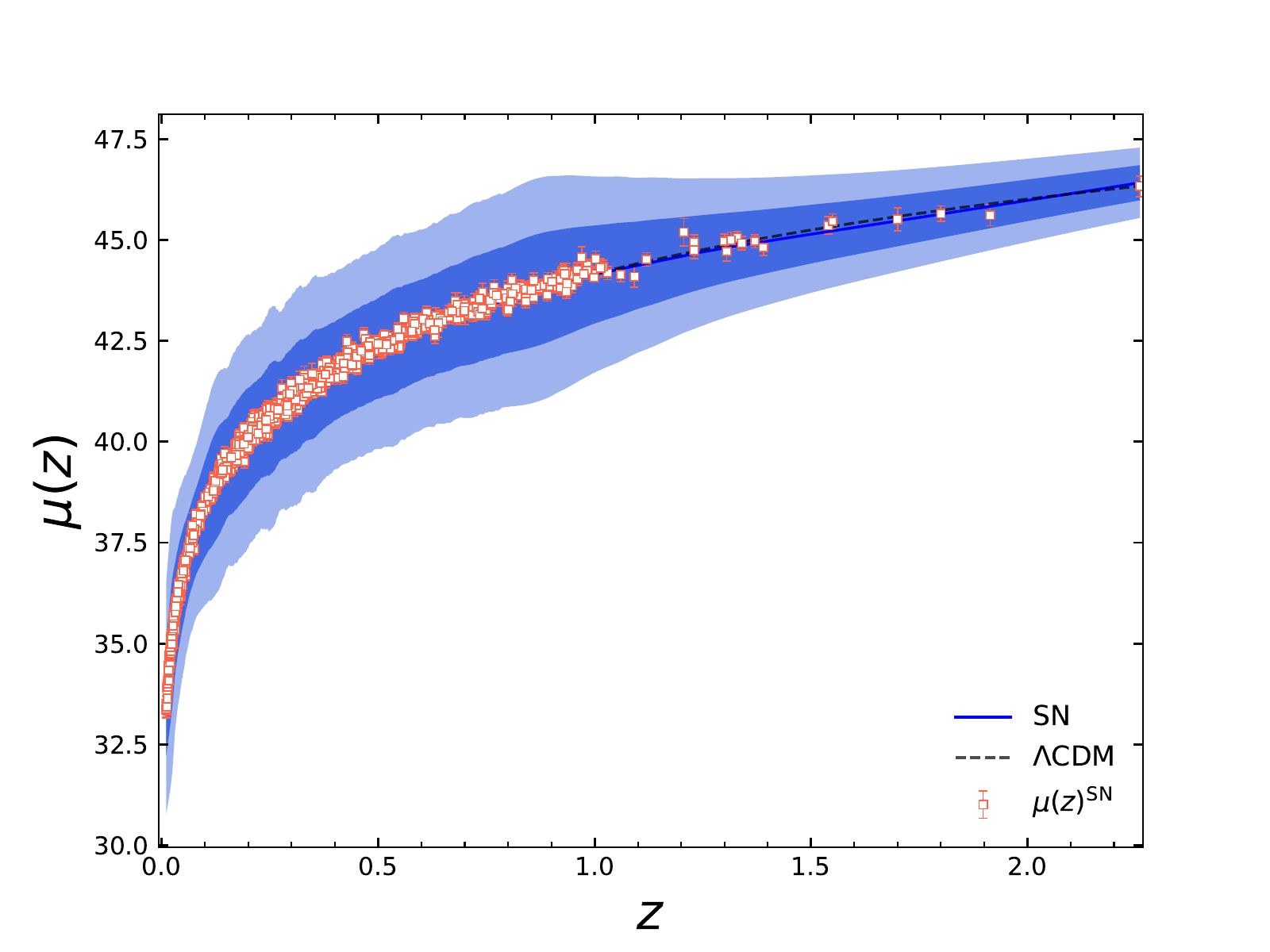}
   \includegraphics[width=0.325\columnwidth]{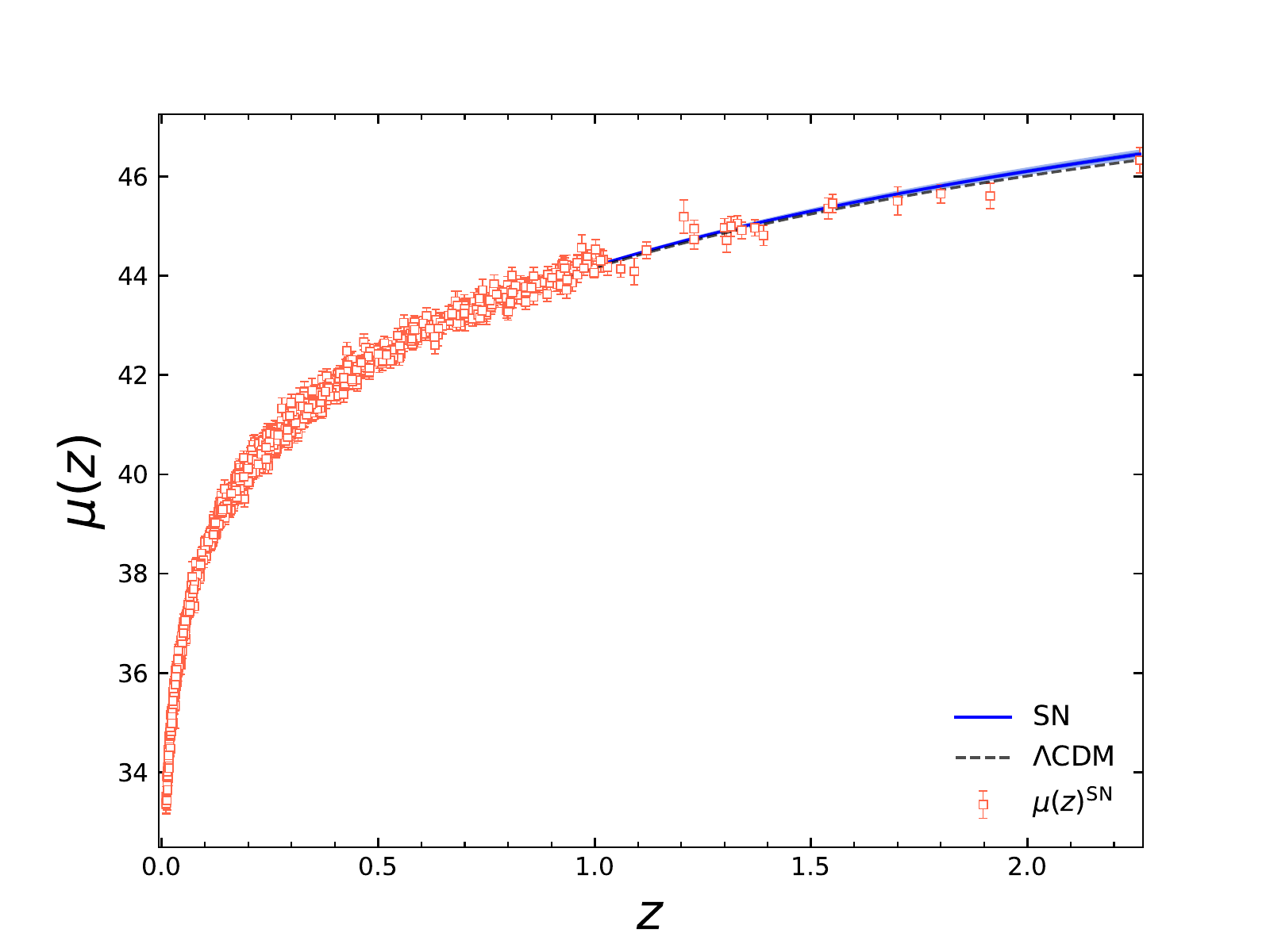}
   \caption{\label{fig:Pantheon_rec}Analysis reconstruction using Pantheon SNeIa data set (orange color points). \textit{Left column:} Cross validation (Eq. (\ref{eq:CV})) is shown for $\alpha$ values ranging between $0$ and $1$. \textit{Middle column:} The LOESS-Simex reconstructed modulus distance $\mu(z)$ diagram is shown (blue color contours) for each prior against a $\Lambda$CDM (PL18) reference (dashed line).  \textit{Right column:} The GP reconstructed modulus distance $\mu(z)$ diagram is shown (blue color contours) for each prior against a $\Lambda$CDM (PL18) reference. In both reconstructions we depict the 1$\sigma$ and 2$\sigma$ confidence regions of the reconstructed data. From \textit{Top} to \textit{Bottom} we denote the $H_0$ prior cases as: TRGB, HW and R19, respectively.}\label{fig:SN_recon}
\end{center}
\end{figure}

\begin{figure}[H]
\begin{center}
    \includegraphics[width=0.325\columnwidth]{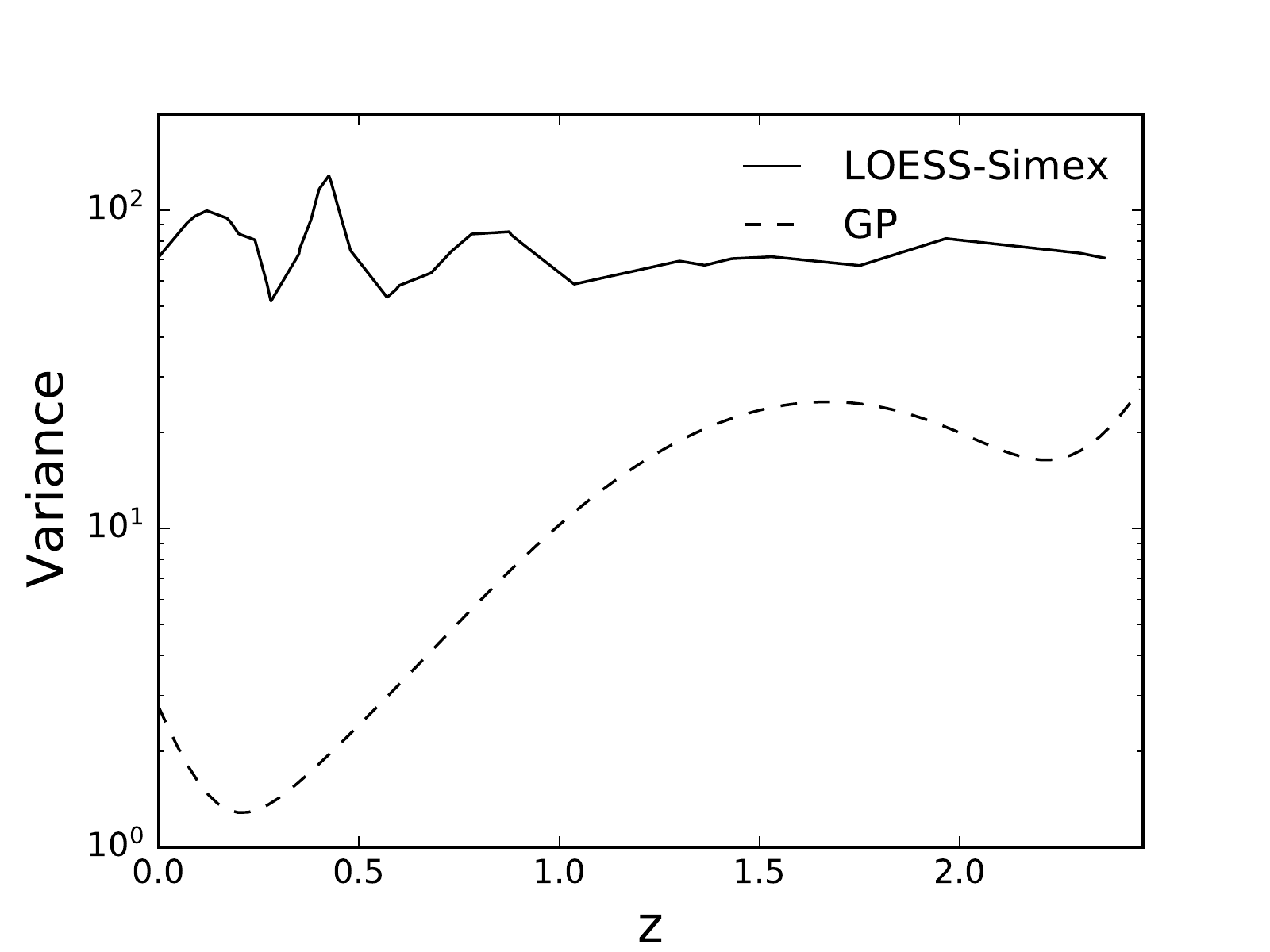}
    \includegraphics[width=0.325\columnwidth]{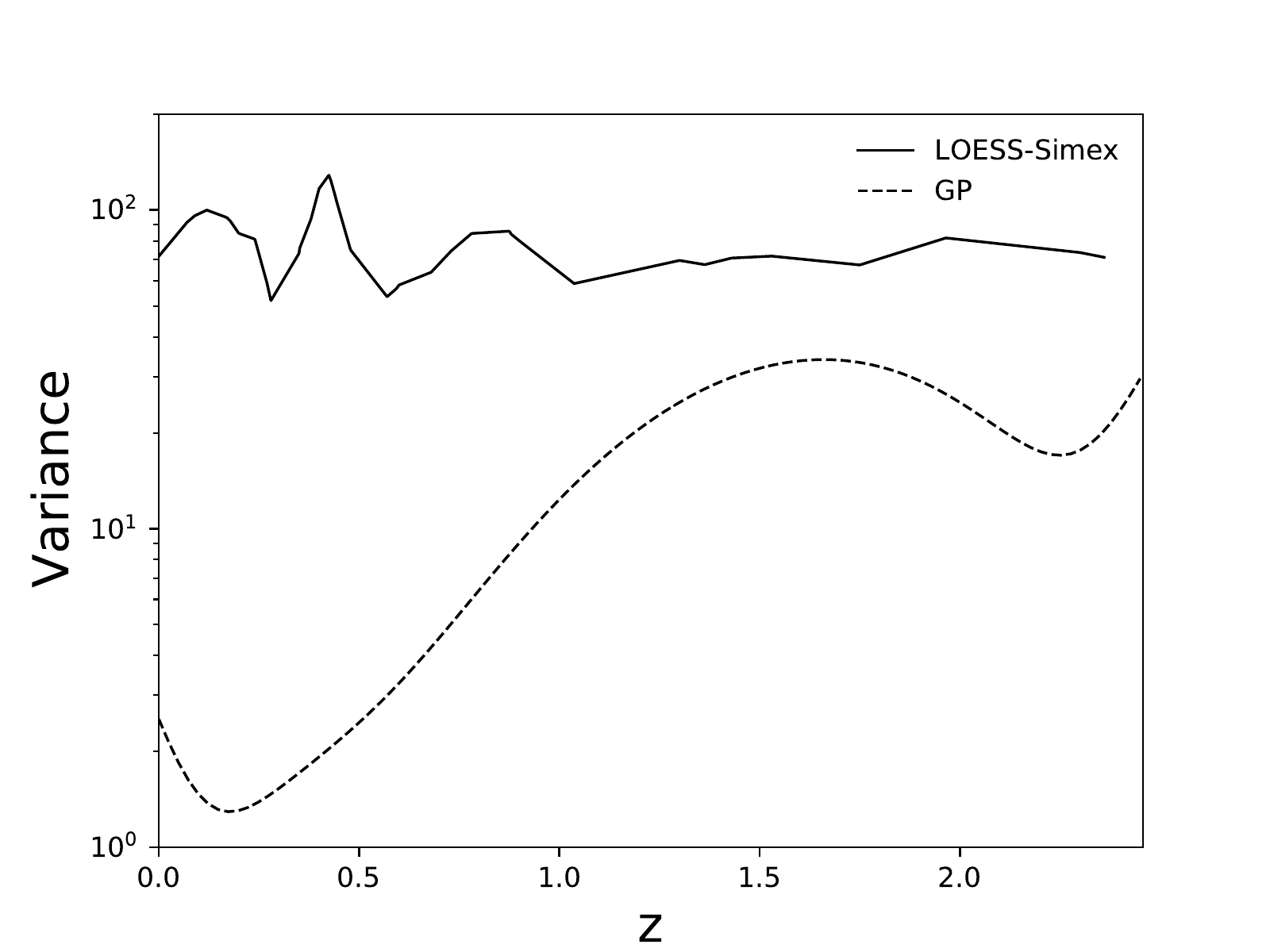}
    \includegraphics[width=0.325\columnwidth]{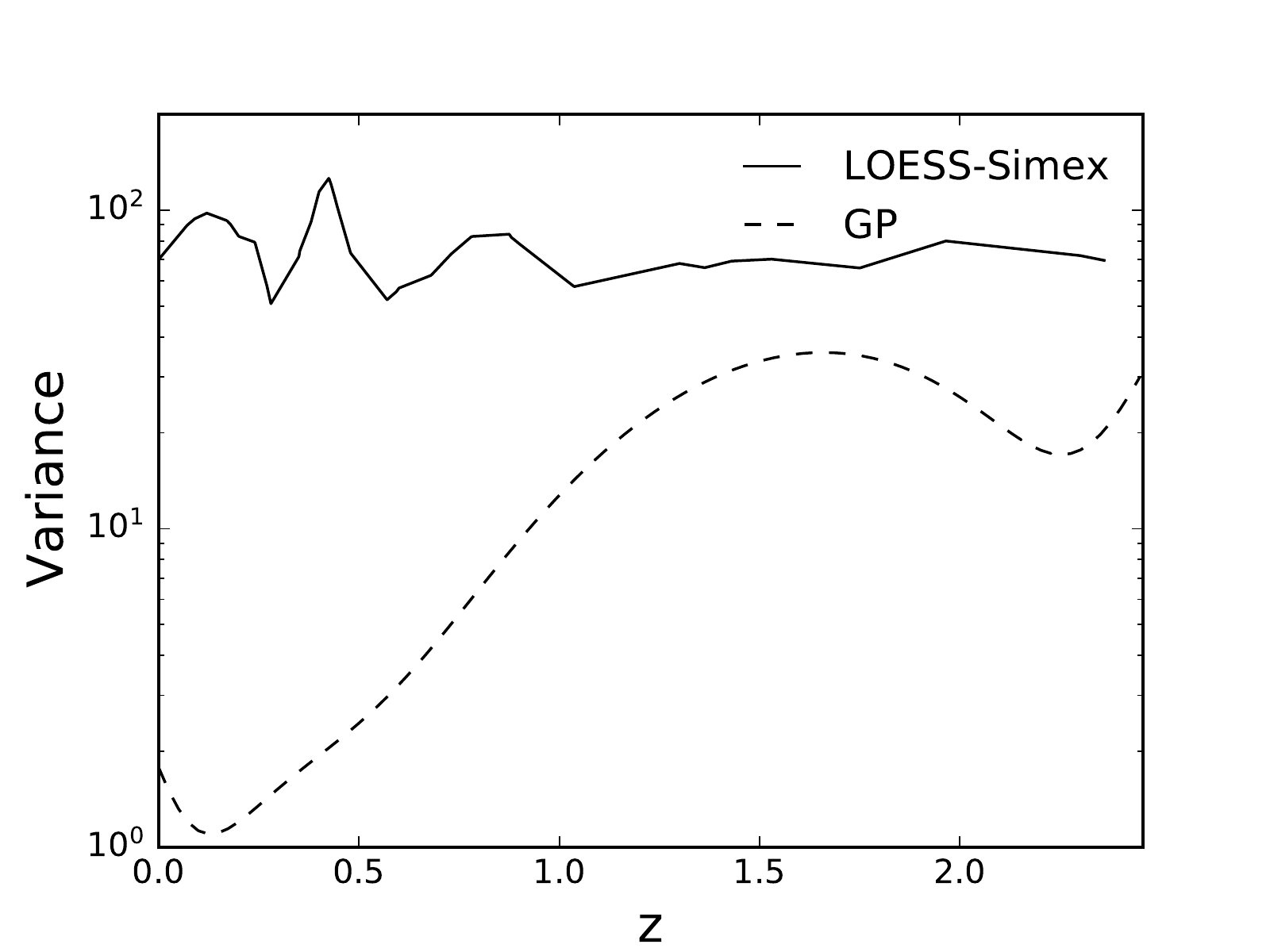}
  \includegraphics[width=0.325\columnwidth]{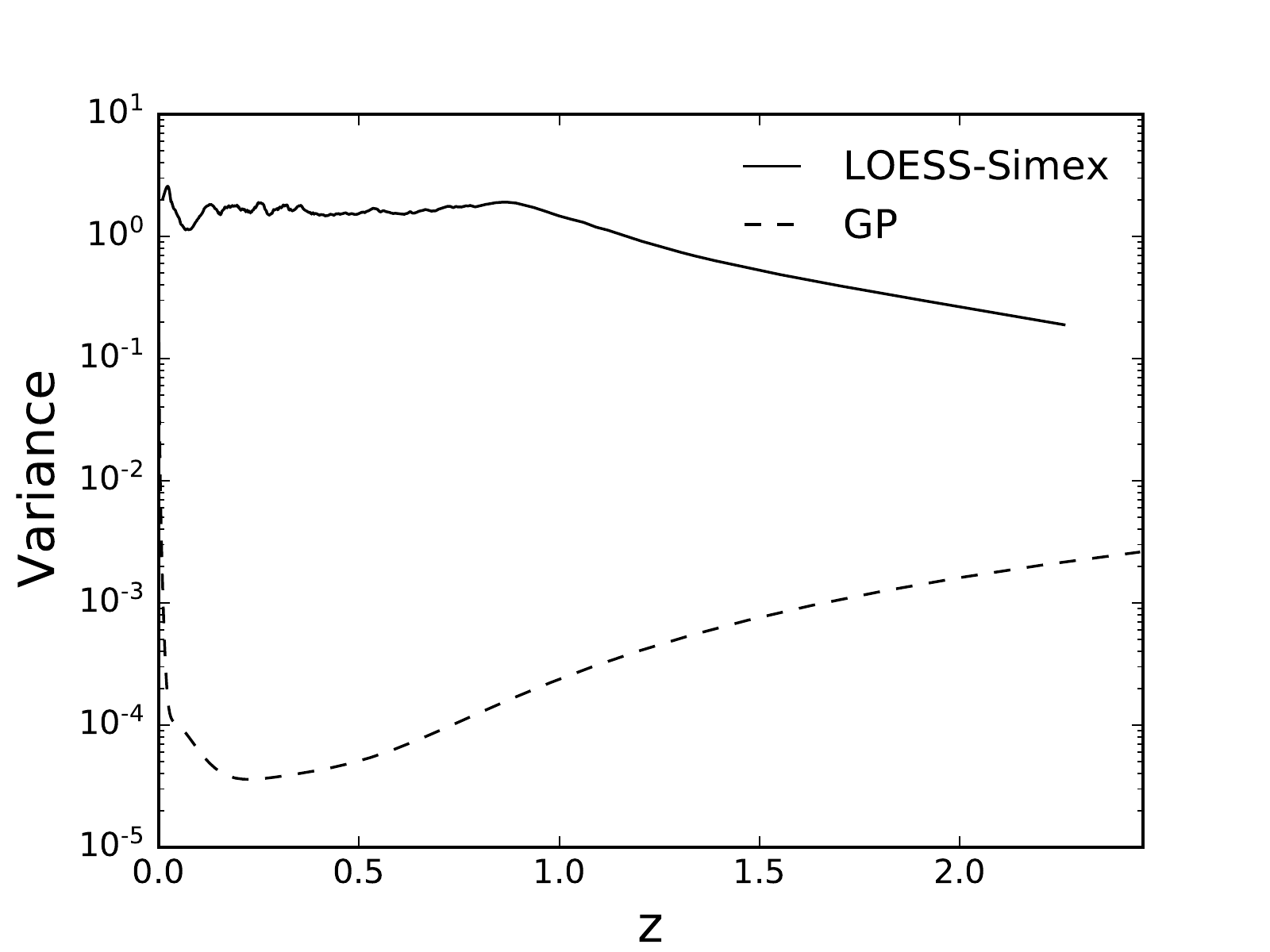}
  \includegraphics[width=0.325\columnwidth]{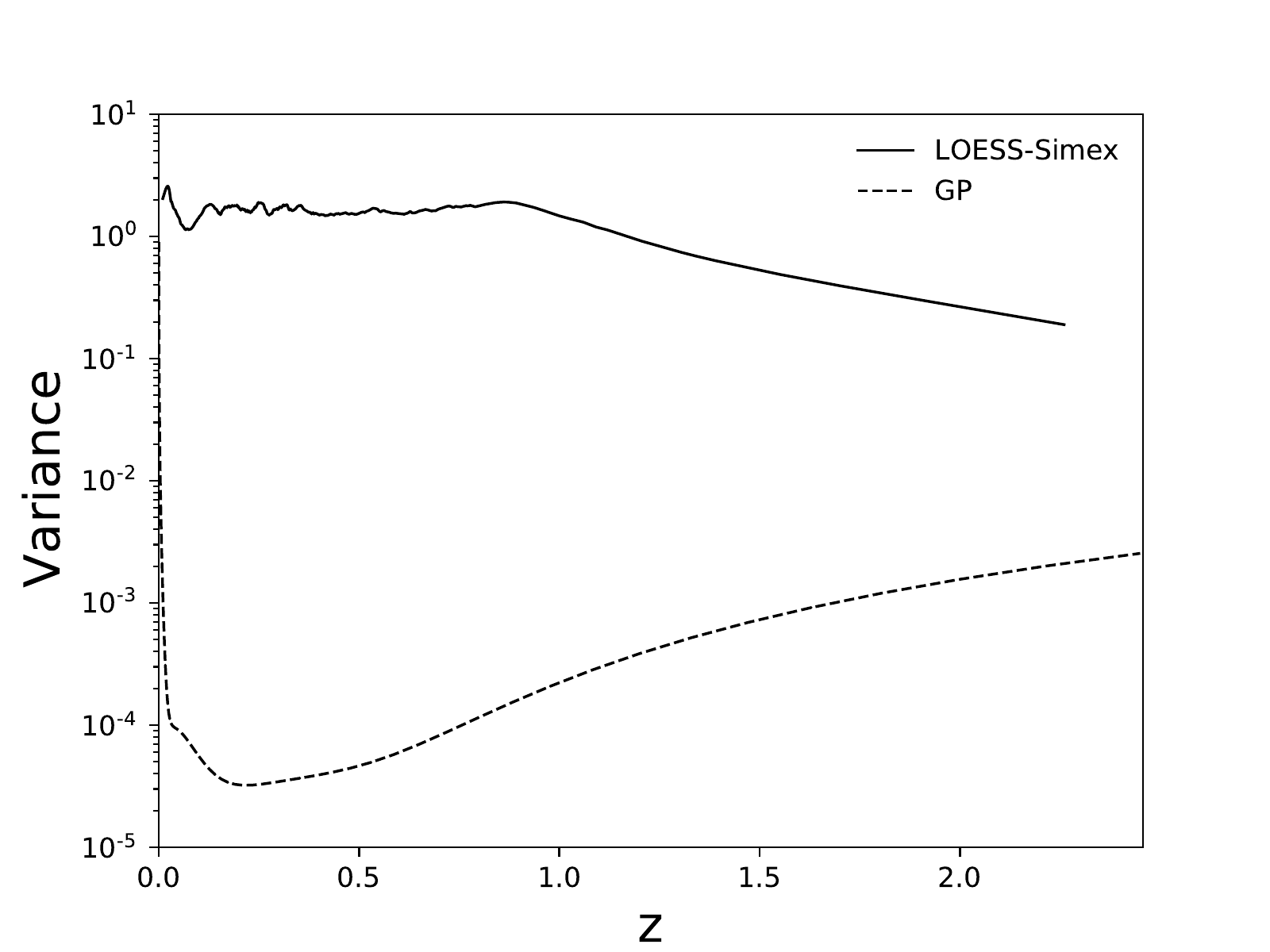}
  \includegraphics[width=0.325\columnwidth]{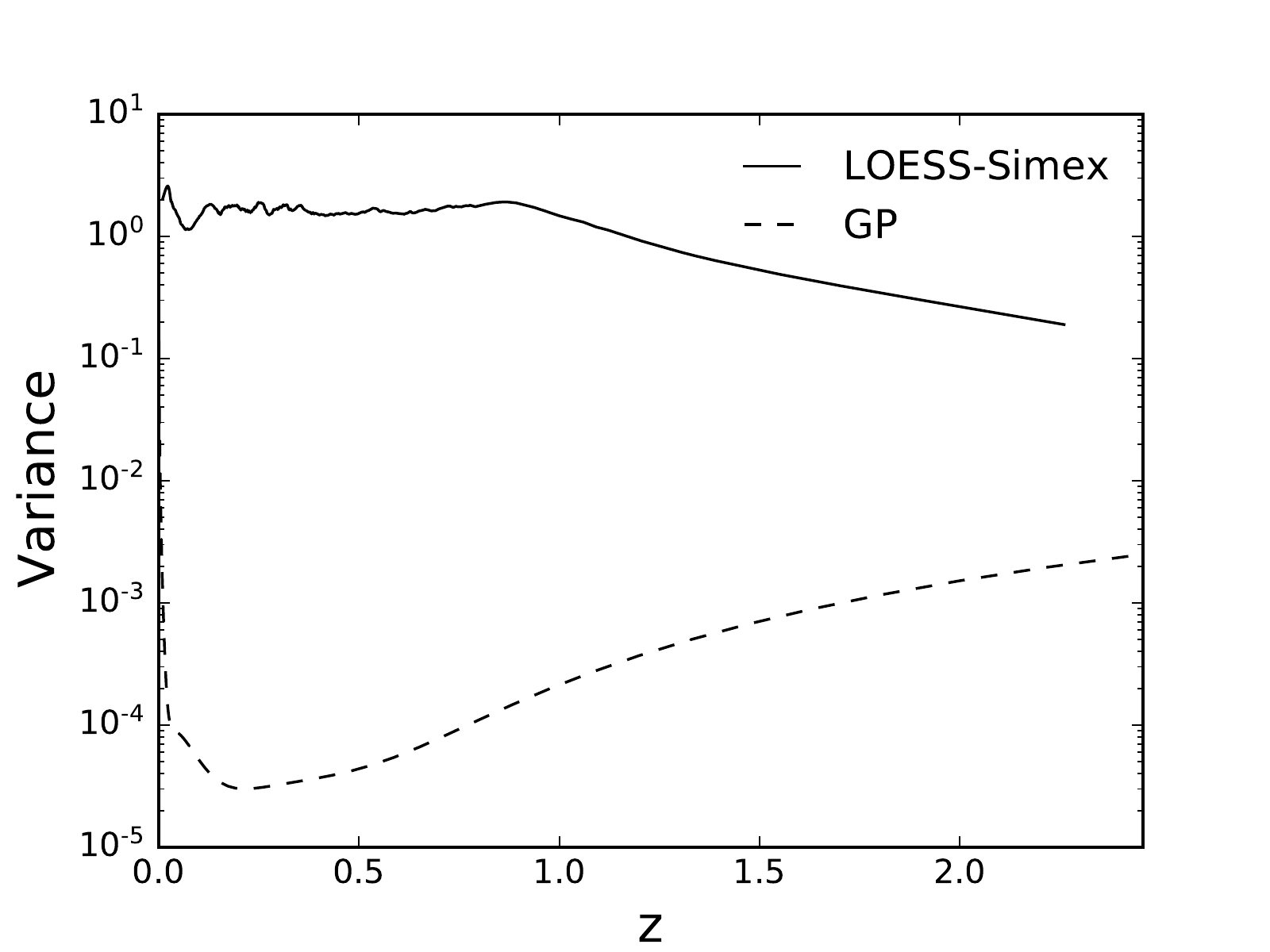}
   \caption{\label{fig:Compar}{Top: Comparisons of the variance are shown for the three priors for the CC+BAO data set.
   Bottom: Comparisons of the variance are shown for the three priors for the Pantheon data set. From \textit{Left} to \textit{Right} we denote the $H_0$ prior cases as: TRGB, HW and R19, respectively.}}
\end{center}
\end{figure}

To this end, the reconstructions for the CC+BAO data set combination shown in Fig.~\ref{fig:CC_BAO_recon} as well as the Pantheon data set reconstructions shown in Fig.~\ref{fig:SN_recon} show good agreement between the mean reconstructed functions and their respective observational data for each method and prior combination. However, the respective uncertainties between both approaches is the most drastic difference between the methods. In this context, Fig.~\ref{fig:Compar} shows the discrepancy between the variances for reconstructions using both the CC+BAO and SNeIa data sets in the background of all three priors. Here, we observe that the selection of priors has little to no effect on the different variances that both reconstructions result in.

Another important property to notice in the CC+BAO combination is that LOESS-Simex retains a variance that largely remains at the same order of magnitude while the GP reconstruction has a dependence on the density of actual observational data points in any particular redshift region. This is to be expected since LOESS-Simex reconstructs data points with an equal contribution from each point in the data set while GP is impacted by the density of points between different regions on the redshift reconstruction space. We also notice that GP predicts much lower uncertainties as compared with LOESS-Simex which is a result of the way in which it assumes the data points to be sourced, namely that each point is a peak in a background of a normally distributed set of data points. Thus, GP makes stronger assumptions on the underlying data being probed. Finally, despite the impact of the prior being minor, higher $H_0$ priors do lead to slightly lower variances across the redshift range being considered with the biggest effect being sourced by the R19 prior since this has the highest prior value. Precisely the same overall behaviour is observed for the Pantheon reconstructed variance shown in Fig.~\ref{fig:Compar} in which the fluctuations in the GP variances is less pronounced.

\begin{figure}[H]
\begin{center}
    \includegraphics[width=0.325\columnwidth]{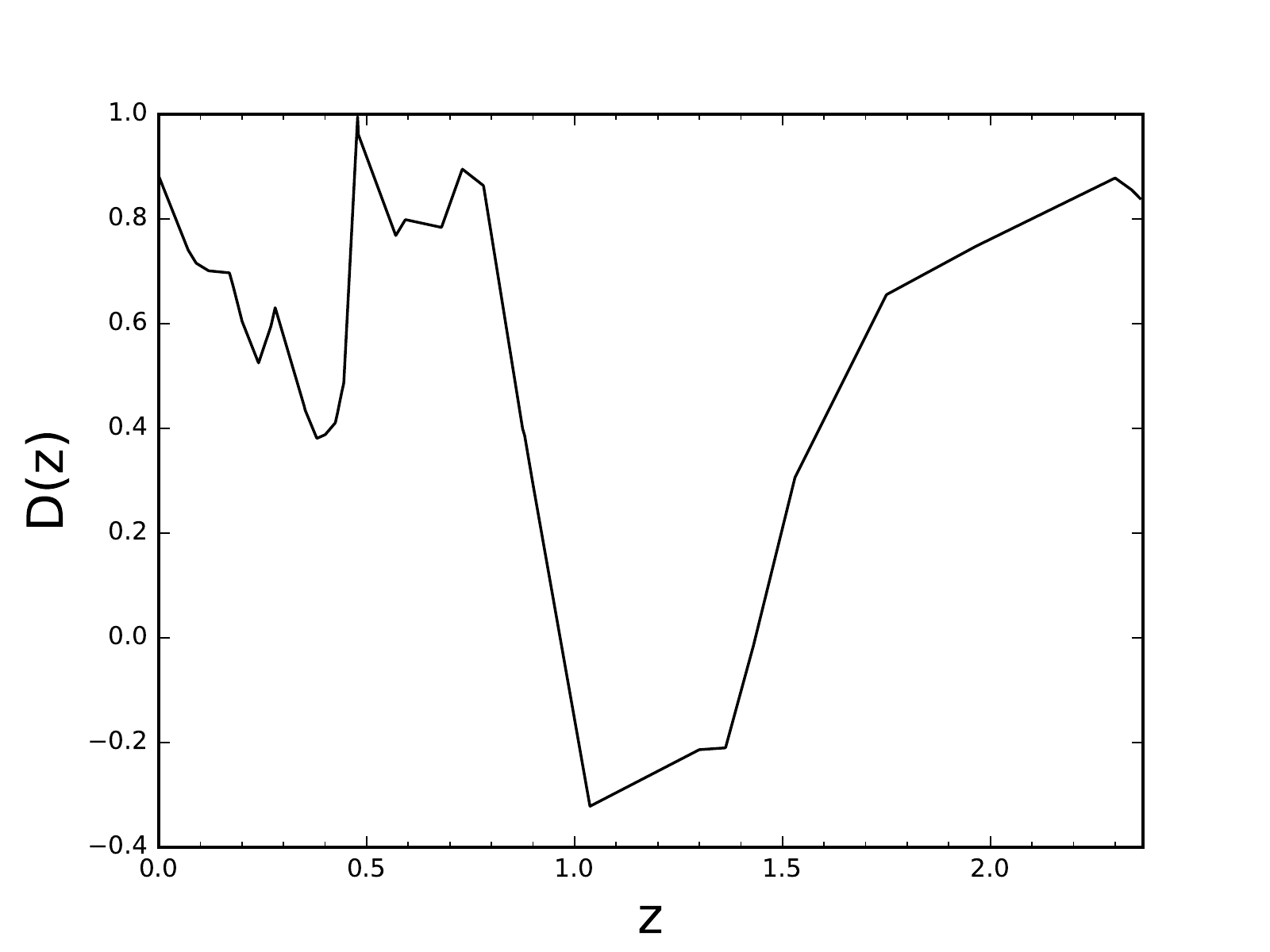}
    \includegraphics[width=0.325\columnwidth]{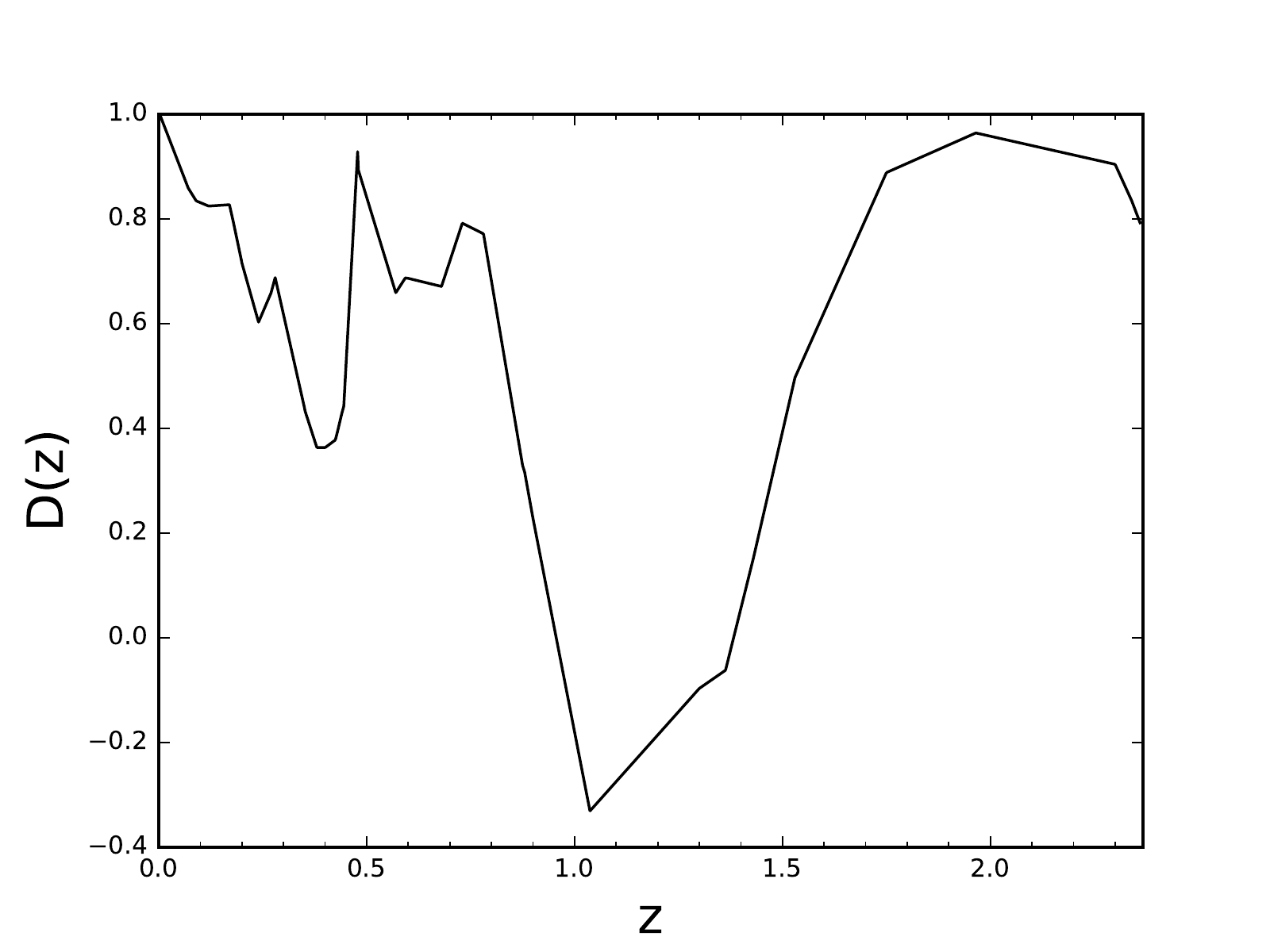}
    \includegraphics[width=0.325\columnwidth]{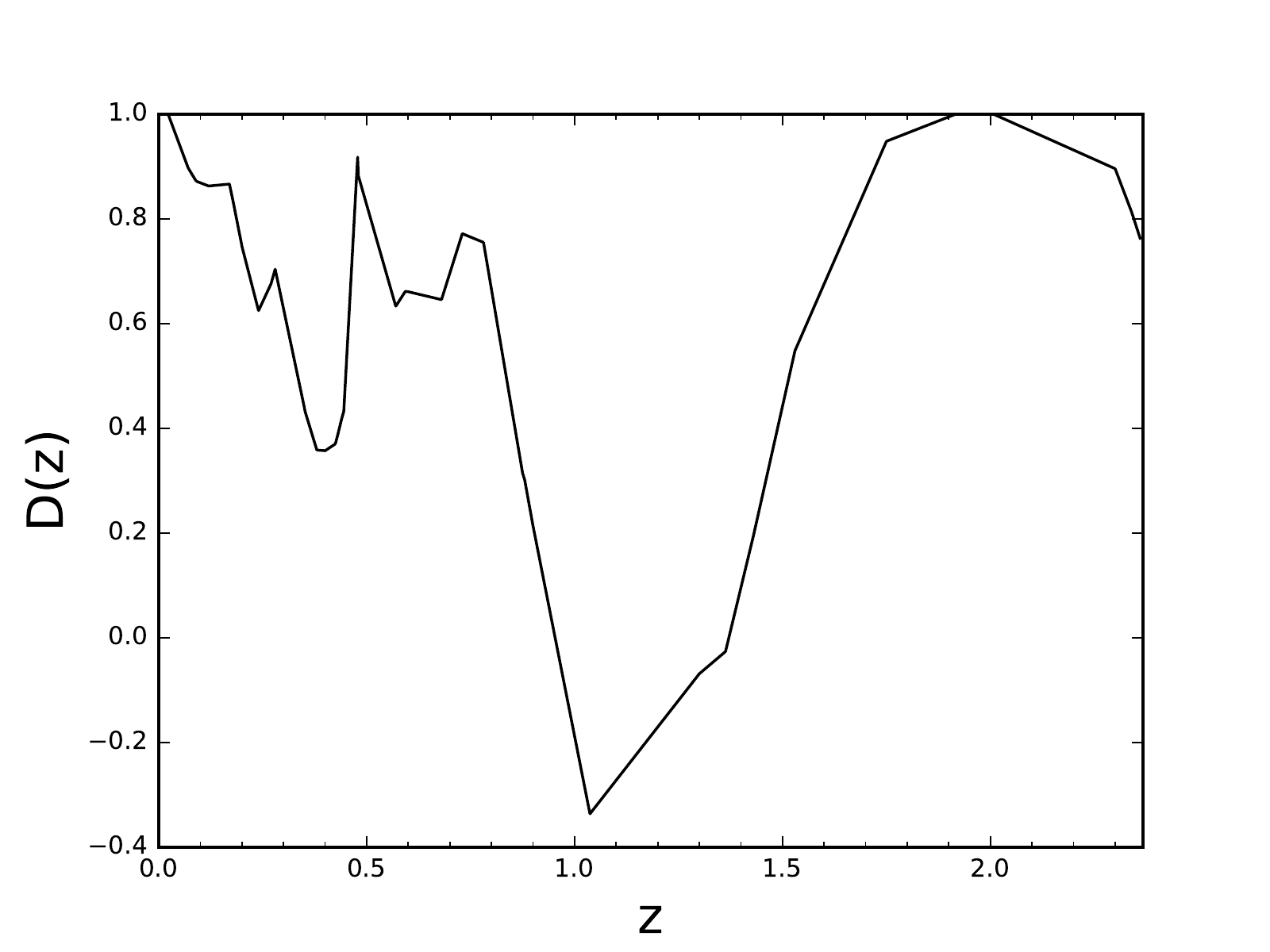}
 \includegraphics[width=0.325\columnwidth]{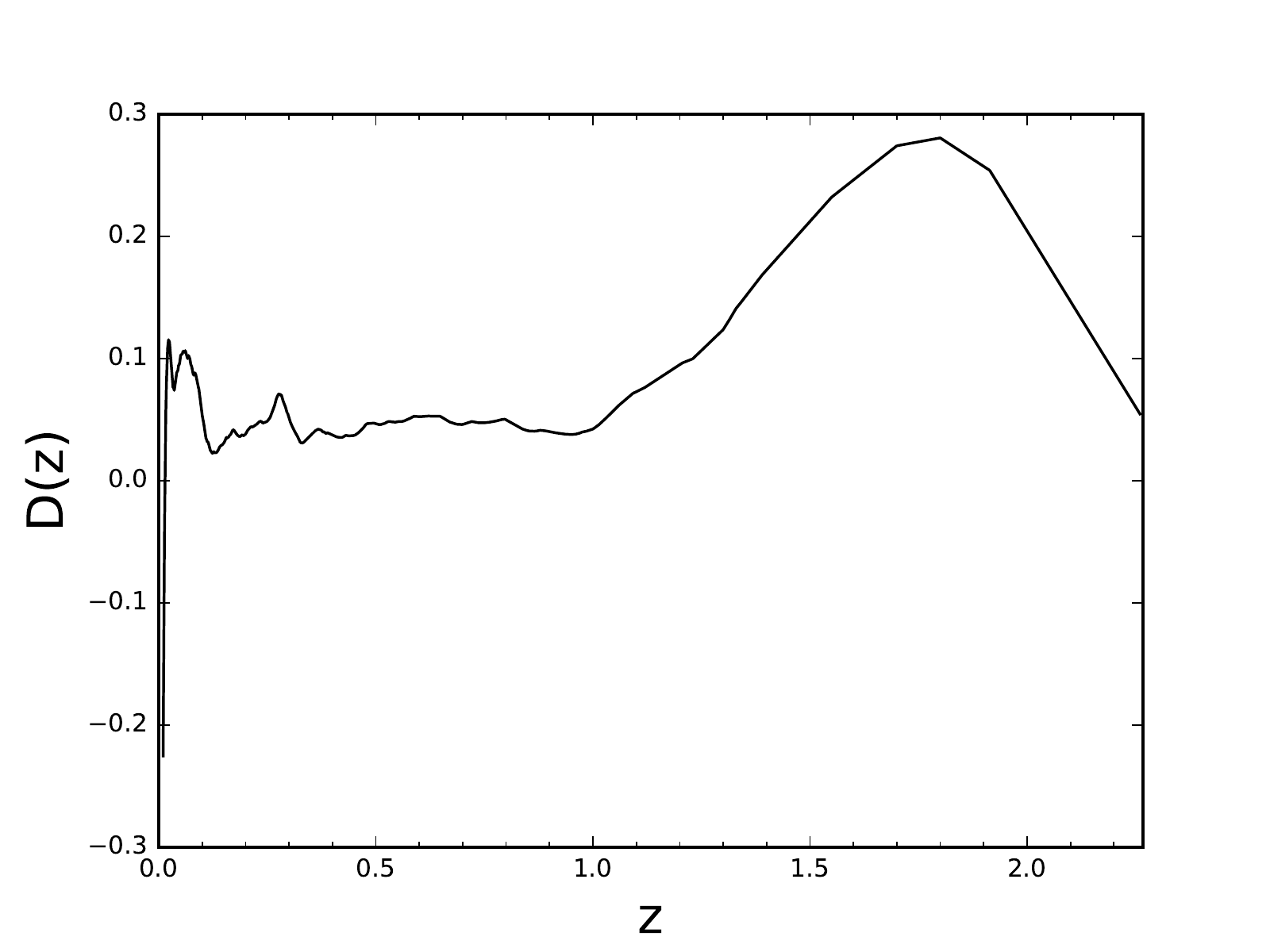}
 \includegraphics[width=0.325\columnwidth]{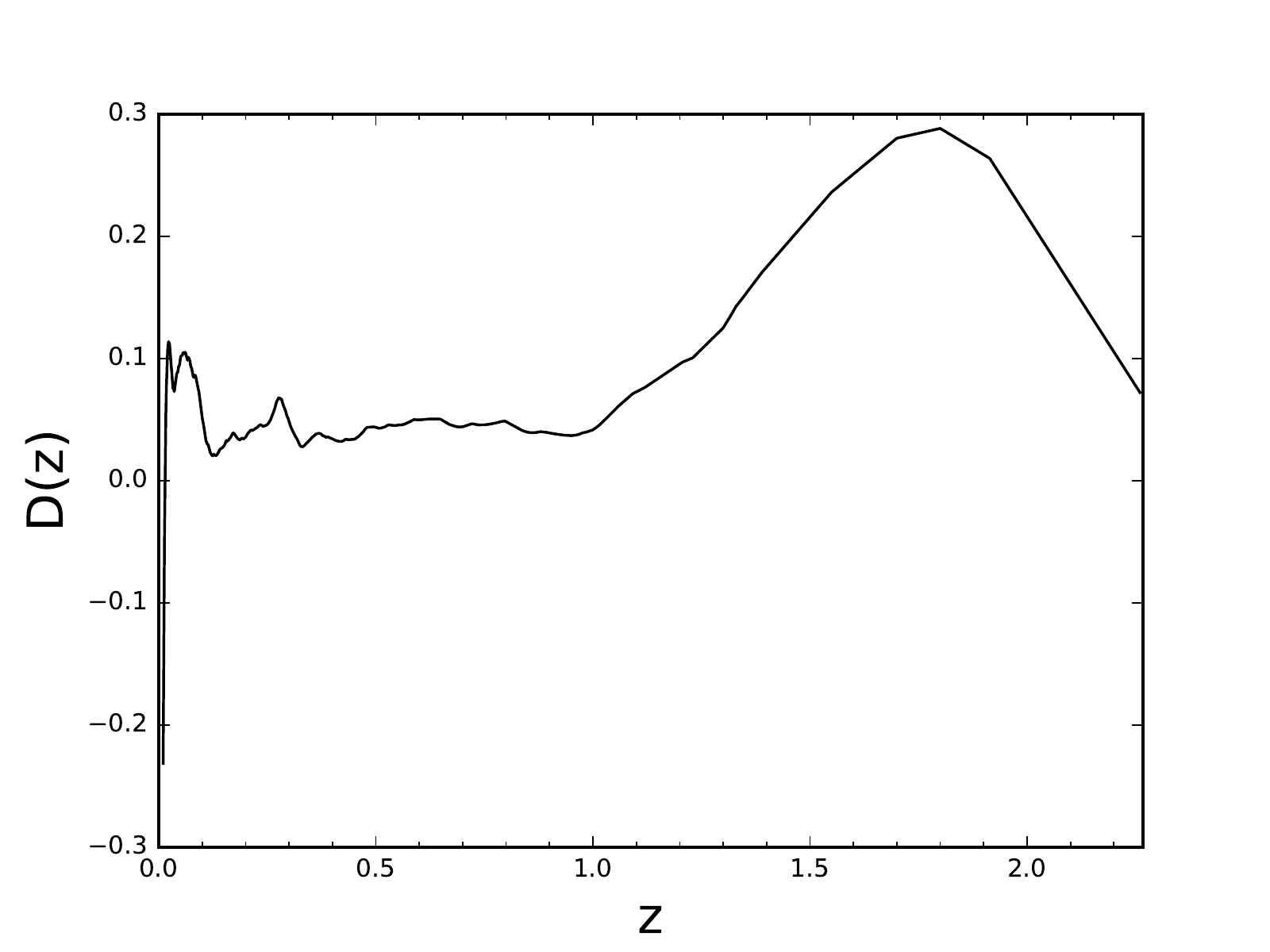}
 \includegraphics[width=0.325\columnwidth]{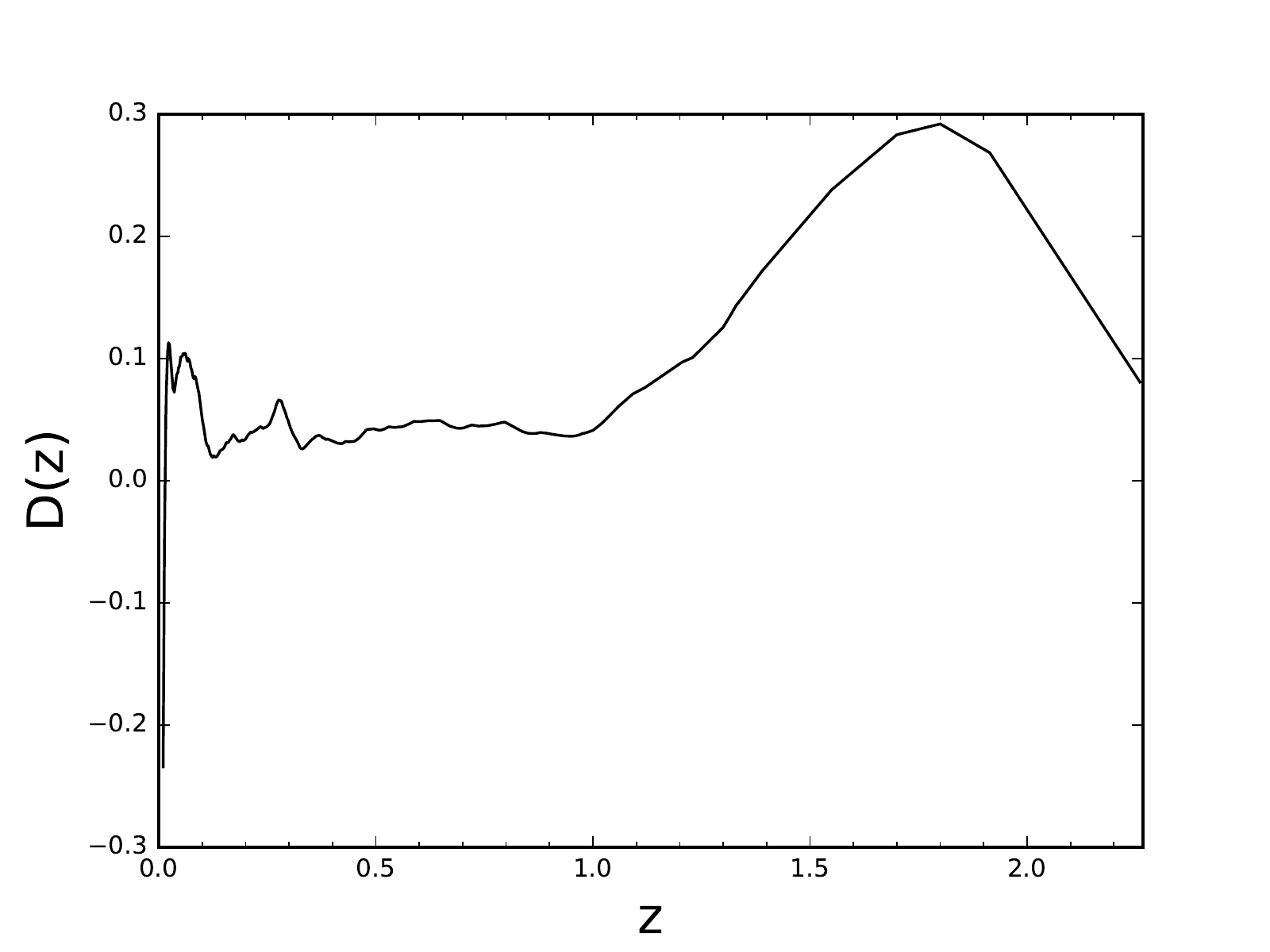}
   \caption{\label{sigma_distance}{Top: The distance between the two reconstructions is shown in $\sigma$ units for the CC+BAO data set.
   Bottom: The distance between the two reconstructions is shown in $\sigma$ units for the Pantheon data set. From \textit{Left} to \textit{Right} we denote the $H_0$ prior cases as: TRGB, HW and R19, respectively.}}
\end{center}
\end{figure}

Both reconstruction approaches can be compared directly by determining the distance between their predicted reconstruction values (in units of $\sigma$), which can be defined as
\begin{eqnarray}
    D(z)& =& \frac{Q_{\rm GP}(z) - Q_{\rm LS}(z)}{\sqrt{\sigma_{\rm GP}^2(z) + \sigma_{\rm LS}^2(z)}}\,,
\end{eqnarray}
where $Q$ denotes the Hubble parameter $H(z)$ for the CC+BAO combination, and $\mu(z)$ for the Pantheon SNeIa data set. We show the evolution of the distance parameter $D(z)$ for the CC+BAO and Pantheon data sets in Fig.~\ref{sigma_distance}. In both cases the reconstructions never cross the 1$\sigma$ line and in the Pantheon reconstruction, this limit is much stricter which is a result of the much larger quantity of data points. As in Fig.~\ref{sigma_distance}, the effect of the priors has little impact on the general behavior of $D(z)$. On the other hand, there is a weak correlation between higher distances at higher redshifts for high priors such as R19. For the CC+BAO data set combination, GP predicts larger values for the Hubble parameter except for a small neighbourhood at about $z = 1.2$ (mostly prior independent). In the Pantheon data set case, the LOESS-Simex reconstruction surpasses the GP predicted values only in the initial neighbourhood of $z=0$ (also prior independent). However, in this second case, the predicted values are much closer together and, generally, much closer to the data points themselves.

Furthermore, the value measured by the following $d(z)$ indicator quantifies the displacement of both reconstructions against the original data. However, to quantify the quality of the entire reconstructions, we sum its square through
\begin{equation}
    d(z) = \frac{Q_{\rm reconstructed}(z) - Q_{\rm observation}(z)}{\sqrt{\sigma_{\rm reconstructed}^2(z) + \sigma_{\rm observation}^2(z)}}\,,
\end{equation}
which gives another important measure of the approximation of either approach to producing the original data (in units of $\sigma$). We denote the sum as
\begin{equation}
    \Upsilon = \sum_{z} d^2(z)\,,
\end{equation}
which is used to produce Table~\ref{table:comparative_comp}.

\begin{table}[t]
\centering
\setlength\extrarowheight{2pt}
\begin{tabular}{ c c c c  }
    \hline
    Prior & Reconstruction method & $\Upsilon_{\rm CC+BAO}$ & $\Upsilon_{\rm SN}$ \\
    \hline
    \multirow{2}{*}{TRGB} & GP & 16.53 & 826.41 \\
     & LS & 16.68 & 16.60 \\
    \hline
    \multirow{2}{*}{HW} & GP & 15.72 & 917.93 \\
     & LS & 16.84 & 16.35 \\
    \hline
    \multirow{2}{*}{R19} & GP & 15.66 & 976.11 \\
     & LS & 16.86 & 16.22 \\
    \hline
\end{tabular}
\caption{Values of $\Upsilon$ are shown for all combinations of reconstruction methods and priors for the CC+BAO and SNeIa data sets.}
\label{table:comparative_comp}
\end{table}

In Table \ref{table:comparative_comp}, the value of $\Upsilon$ is shown for each combination of prior $H(z)$ value together with both reconstructions. In the background of this setting, both reconstructions perform at a similar order of magnitude with the GP reconstruction producing values that systematically have smaller variances. Table \ref{table:comparative_comp} also shows the over-fitting by the GP reconstruction, which in this instance produces very small variances which thus produce large values of $\Upsilon$.


\section{Conclusions}
\label{sec:conclusions}

In the current era of precision cosmology, the amount of observational data is significantly increasing. Moreover, the robustness of cosmological data is always being enhanced due to state-of-the-art cosmological probes of the Universe. Therefore it is natural to adopt machine learning techniques to cosmological data in order to infer model independent results which are purely driven by observational data. In this analysis we focused on the model independent reconstruction of functions via two distinct machine learning techniques, namely LOESS-Simex and GP which are respectively introduced in Sec. \ref{sec:Loess_intro} and Sec. \ref{sec:GP_intro}.

As described in Sec. \ref{sec:CC_BAO_data} and Sec. \ref{sec:SN_data}, we here considered the model independent reconstruction of $H(z)$ by using the CC and BAO data, as well as the reconstruction of $\mu(z)$ by adopting the Pantheon SNeIa data. We further considered the latter data sets in the presence of the TRGB, HW and R19 $H_0$ priors. In the Pantheon SNeIa data set, these prior values were introduced via the calibration of the original SNeIa data set, as described in Sec. \ref{sec:SN_data}. In our study, we considered observational data directly to avoid convoluting our results with that of calibration, for instance. A further study may better incorporate this together with a fuller understanding of the tension in the value of the $H_0$ constant, among other effects such as the tendency of BAO data to prefer a lower value of this parameter.

Further to our GP and LOESS-Simex reconstructions of $H(z)$ and $\mu(z)$, we statistically compared the two methods in Sec. \ref{sec:comparative}. In general, both machine learning techniques were able to reconstruct $H(z)$ and $\mu(z)$ data very well. Moreover, the $\Lambda$CDM model was found to be in very good agreement with all the inferred GP and LOESS-Simex reconstructions. We observed that the variance in GP reconstructions is smaller than that of LOESS-Simex, although in the case of LOESS-Simex the variance does not fluctuate as much as in the GP scenarios. Generically, the LOESS-Simex technique is characterised by a broader and more conservative uncertainty region than that of GP. We also showed that these distinguishing features of the considered data driven reconstruction techniques are independent from the adopted $H_0$ prior values. 

Undoubtedly, machine learning techniques are already giving new insights to the theoretical cosmological framework via a more elaborate exploration and interpretation of observational data. Therefore, further development and implementation of machine learning techniques to cosmological data would facilitate our search for specific cosmological signatures, which would shed light on the set of viable cosmological models.


\begin{acknowledgments}
CE-R is supported by the Royal Astronomical Society as FRAS 10147 and by DGAPA-PAPIIT-UNAM Project IA100220. This article is based upon work from CANTATA COST (European Cooperation in Science and Technology) action CA15117, EU Framework Programme Horizon 2020. The authors would like to acknowledge networking support by the COST Action CA18108 and funding support from Cosmology@MALTA which is supported by the University of Malta. We are grateful to Reginald Christian Bernardo for important feedback on new methods in GP regression. This research has been carried out using computational facilities procured through the European Regional Development Fund, Project No. ERDF-080 ``A supercomputing laboratory for the University of Malta''.
\end{acknowledgments}


\appendix
\section{GP subtleties} \label{appendix:GP}
\begin{figure}[t!]
    \centering
    \includegraphics[width=0.45\columnwidth]{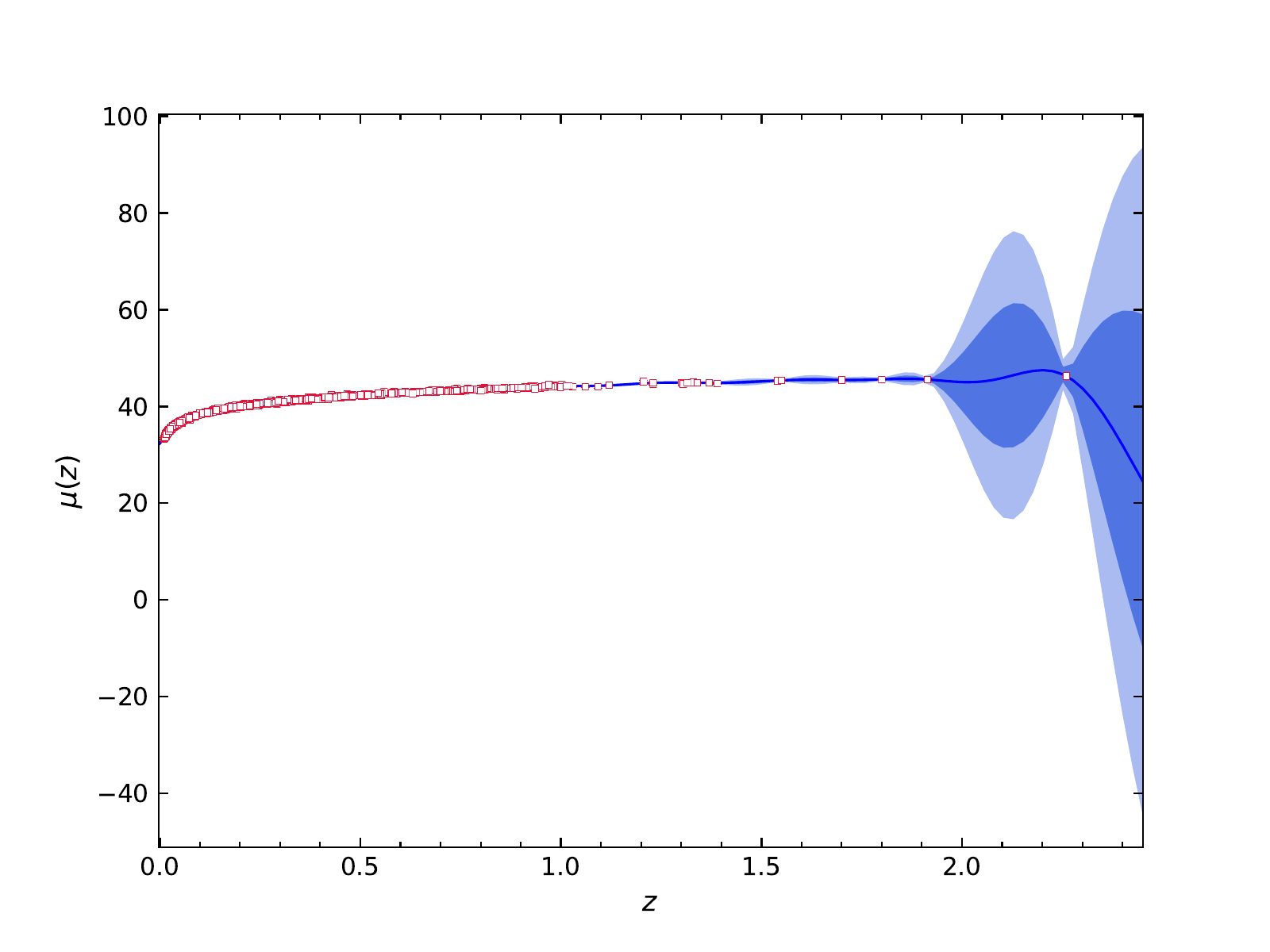}
    \includegraphics[width=0.45\columnwidth]{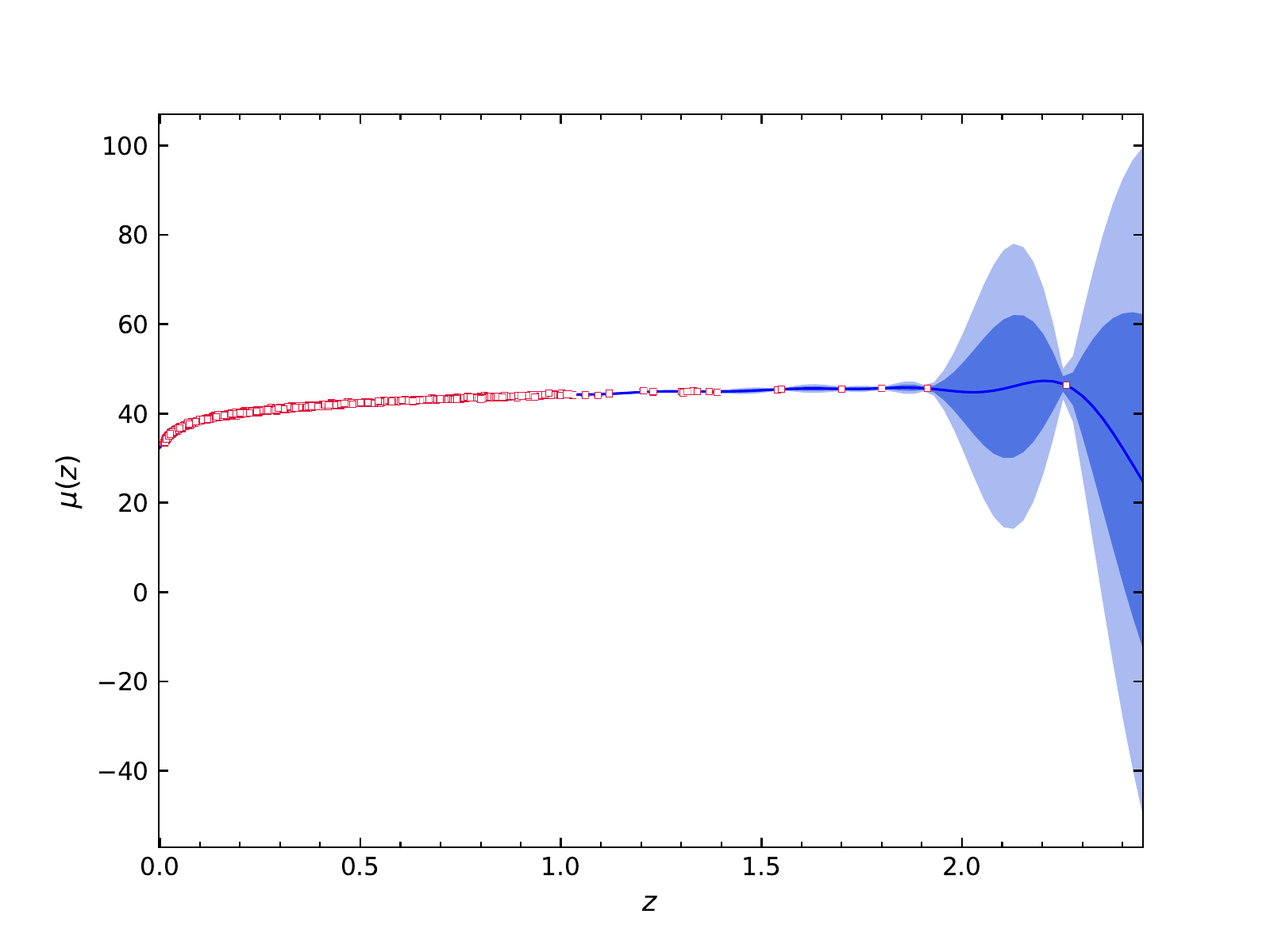}
    \includegraphics[width=0.45\columnwidth]{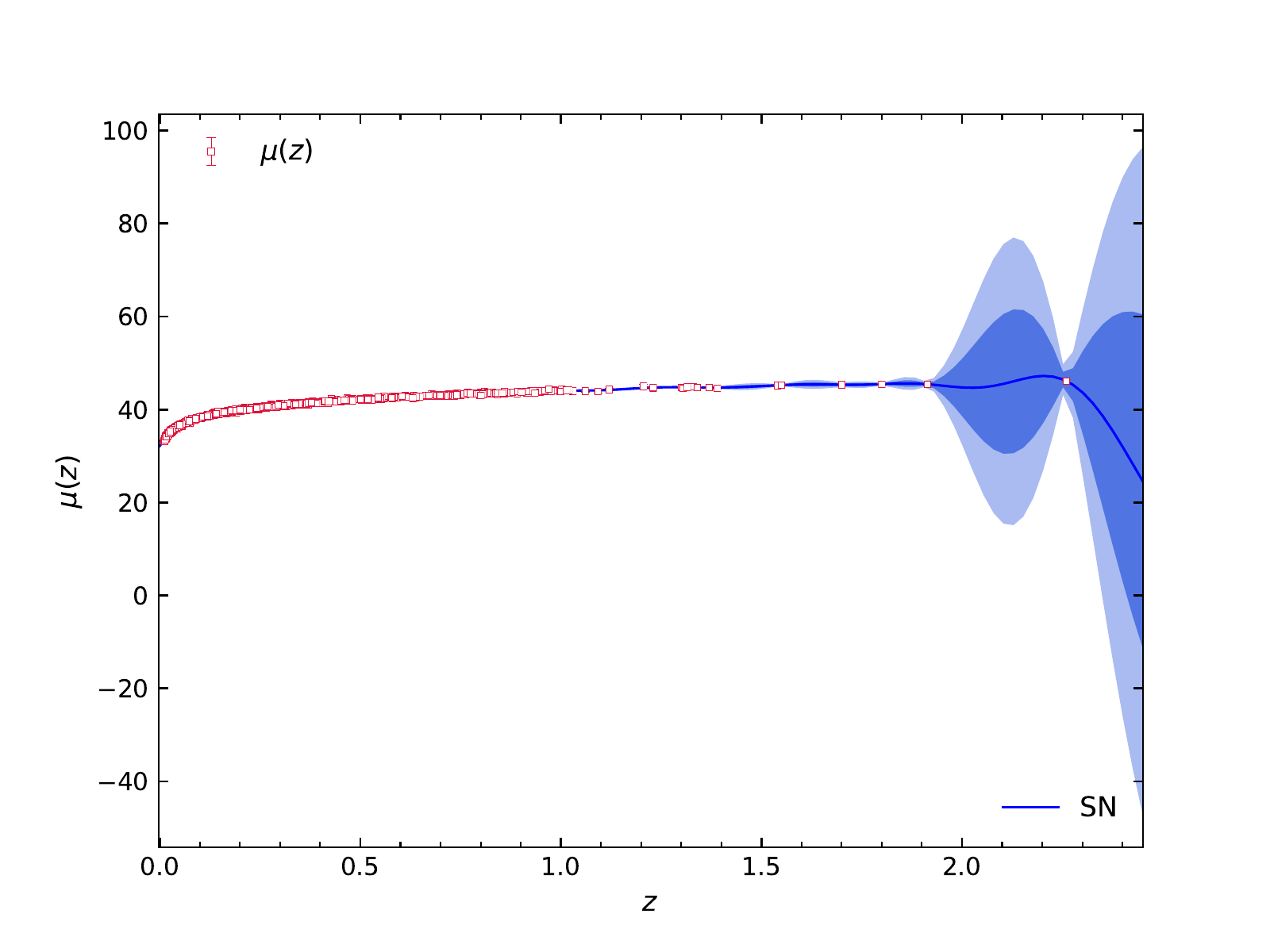}
    \caption{GP reconstructions of the Pantheon SNeIa data set using a linear redshift scale with the TRGB (top left), R19 (top right) and HW (bottom) $H_0$ prior values.}
    \label{fig:mu_linear}
\end{figure}

We would like to note that the GP technique and the GaPP code are sensitive to the chosen horizontal axis scale of the data points, and hence to the data set that one aims to reconstruct. Indeed, in Fig. \ref{fig:Pantheon_rec} we have adopted a logarithmic horizontal scale for the data points, which led to very tight constraints on the reconstructed function of $\mu(z)$. However, when we considered a linear horizontal scale for the Pantheon SNeIa data set, we inferred a more conservative and less robust reconstruction, as illustrated in Fig. \ref{fig:mu_linear}. In the case of a linear redshift scale, the GP hyperparameters are dominated by the low redshift data points. Consequently, the derived hyperparameter values do not fit the high redshift data points very well in this case. Indeed, as illustrated in Fig. \ref{fig:mu_linear}, at high redshifts, the fitted hyperparameters do not lead to a continuously and monotonically smooth reconstruction band. On the other hand, when a logarithmic redshift scale is considered, the hyperparameters fit the entire data set very well, as illustrated in Fig. \ref{fig:Pantheon_rec}. This observation could be explained via the inferred GP kernel hyperparameter values of $\sigma_f$ and $l_f$, as reported in Table \ref{table:hyperparameters_comp_SN}. It is clear that the optimal value of $l_f$ is significantly altered by the chosen redshift scale of the data points, which is expected since the length separation from one data point to another is altered when rescaling from a linear to a logarithmic axis.

\begin{table}[h]
\begin{center}
\setlength\extrarowheight{2pt}
\begin{tabular}{c c c c}
    \hline
    Prior & Scale & $\sigma_f$ & $l_f$ \\
    \hline
    \multirow{2}{*}{TRGB}& Linear & 42.89 & 0.18 \\
    &Logarithmic & 61.32 & 7.36 \\
    \hline
    \multirow{2}{*}{HW}& Linear & 44.61 & 0.18\\
    &Logarithmic & 59.37 & 6.98\\
    \hline
    \multirow{2}{*}{R19}& Linear & 46.55 & 0.18\\
    &Logarithmic & 58.76 & 6.82\\
    \hline
\end{tabular}
\end{center}
\caption{A comparison of the derived hyperparameter values of $\sigma_f$ and $l_f$ with the TRGB, HW and R19 Hubble constant priors when adopting a linear or a logarithmic redshift scale in the corresponding GP reconstructions.}
\label{table:hyperparameters_comp_SN}
\end{table}

\bibliographystyle{JHEP}
\bibliography{references}

\providecommand{\href}[2]{#2}\begingroup\raggedright\begin{thebibliography}{10}

\bibitem{Bernal:2016gxb}
J.~L. Bernal, L.~Verde and A.~G. Riess, \emph{{The trouble with $H_0$}},
  \href{https://doi.org/10.1088/1475-7516/2016/10/019}{\emph{JCAP} {\bfseries
  10} (2016) 019} [\href{https://arxiv.org/abs/1607.05617}{{\ttfamily
  1607.05617}}].

\bibitem{Perlmutter:1998np}
{\scshape Supernova Cosmology Project} collaboration, \emph{{Measurements of
  $\Omega$ and $\Lambda$ from 42 high redshift supernovae}},
  \href{https://doi.org/10.1086/307221}{\emph{Astrophys.J.} {\bfseries 517}
  (1999) 565} [\href{https://arxiv.org/abs/astro-ph/9812133}{{\ttfamily
  astro-ph/9812133}}].

\bibitem{Riess:1998cb}
{\scshape Supernova Search Team} collaboration, \emph{{Observational evidence
  from supernovae for an accelerating universe and a cosmological constant}},
  \href{https://doi.org/10.1086/300499}{\emph{Astron.J.} {\bfseries 116} (1998)
  1009} [\href{https://arxiv.org/abs/astro-ph/9805201}{{\ttfamily
  astro-ph/9805201}}].

\bibitem{RevModPhys.61.1}
S.~Weinberg, \emph{The cosmological constant problem},
  \href{https://doi.org/10.1103/RevModPhys.61.1}{\emph{Rev. Mod. Phys.}
  {\bfseries 61} (1989) 1}.

\bibitem{Bull:2015stt}
P.~Bull et~al., \emph{{Beyond $\Lambda$CDM: Problems, solutions, and the road
  ahead}}, \href{https://doi.org/10.1016/j.dark.2016.02.001}{\emph{Phys. Dark
  Univ.} {\bfseries 12} (2016) 56}
  [\href{https://arxiv.org/abs/1512.05356}{{\ttfamily 1512.05356}}].

\bibitem{Copeland:2006wr}
E.~J. Copeland, M.~Sami and S.~Tsujikawa, \emph{{Dynamics of dark energy}},
  \href{https://doi.org/10.1142/S021827180600942X}{\emph{Int. J. Mod. Phys. D}
  {\bfseries 15} (2006) 1753}
  [\href{https://arxiv.org/abs/hep-th/0603057}{{\ttfamily hep-th/0603057}}].

\bibitem{Capozziello:2011et}
S.~Capozziello and M.~De~Laurentis, \emph{{Extended Theories of Gravity}},
  \href{https://doi.org/10.1016/j.physrep.2011.09.003}{\emph{Phys. Rept.}
  {\bfseries 509} (2011) 167}
  [\href{https://arxiv.org/abs/1108.6266}{{\ttfamily 1108.6266}}].

\bibitem{Clifton:2011jh}
T.~Clifton, P.~G. Ferreira, A.~Padilla and C.~Skordis, \emph{{Modified Gravity
  and Cosmology}},
  \href{https://doi.org/10.1016/j.physrep.2012.01.001}{\emph{Phys. Rept.}
  {\bfseries 513} (2012) 1} [\href{https://arxiv.org/abs/1106.2476}{{\ttfamily
  1106.2476}}].

\bibitem{DiValentino:2020vvd}
E.~Di~Valentino et~al., \emph{{Cosmology Intertwined III: $f \sigma_8$ and
  $S_8$}},  \href{https://arxiv.org/abs/2008.11285}{{\ttfamily 2008.11285}}.

\bibitem{Aghanim:2018eyx}
{\scshape Planck} collaboration, \emph{{Planck 2018 results. VI. Cosmological
  parameters}},
  \href{https://doi.org/10.1051/0004-6361/201833910}{\emph{Astron. Astrophys.}
  {\bfseries 641} (2020) A6}
  [\href{https://arxiv.org/abs/1807.06209}{{\ttfamily 1807.06209}}].

\bibitem{10.5555/1162254}
C.~E. Rasmussen and C.~K.~I. Williams, \emph{Gaussian Processes for Machine
  Learning (Adaptive Computation and Machine Learning)}. The MIT Press, 2005.

\bibitem{10.1145/3390557.3394126}
J.~Surma, \emph{Hacking Machine Learning: Towards The Comprehensive Taxonomy of
  Attacks Against Machine Learning Systems}, ICIAI 2020. Association for
  Computing Machinery, New York, NY, USA, 2020,
  \href{https://doi.org/10.1145/3390557.3394126}{10.1145/3390557.3394126}.

\bibitem{alma991007710432905601}
J.~M{\o}ller, \emph{Spatial Statistics and Computational Methods}, Lecture
  Notes in Statistics, 173. Springer New York, New York, NY, 1st ed. 2003.~ed.,
  2003,
  \href{https://doi.org/10.1007/978-0-387-21811-3}{10.1007/978-0-387-21811-3}.

\bibitem{kiusalaas_2013}
J.~Kiusalaas, \emph{Numerical Methods in Engineering with Python 3}. Cambridge
  University Press, 3~ed., 2013,
  \href{https://doi.org/10.1017/CBO9781139523899}{10.1017/CBO9781139523899}.

\bibitem{Banerjee}
B.~C. S.~Banerjee and A.~Gelfand, \emph{Hierarchical modeling and analysis for
  spatial data}, .

\bibitem{Cleveland}
W.~S. Cleveland, \emph{Robust locally weighted regression and smoothing
  scatterplots},
  \href{https://doi.org/10.1080/01621459.1979.10481038}{\emph{Journal of the
  American Statistical Association} {\bfseries 74} (1979) 829}.

\bibitem{Gomez-Valent:2018hwc}
A.~Gómez-Valent and L.~Amendola, \emph{{$H_0$ from cosmic chronometers and
  Type Ia supernovae, with Gaussian Processes and the novel Weighted Polynomial
  Regression method}},
  \href{https://doi.org/10.1088/1475-7516/2018/04/051}{\emph{JCAP} {\bfseries
  04} (2018) 051} [\href{https://arxiv.org/abs/1802.01505}{{\ttfamily
  1802.01505}}].

\bibitem{Colgain:2021ngq}
E.~Colg\'ain and M.~M. Sheikh-Jabbari, \emph{{Elucidating cosmological model
  dependence with $H_0$}},  \href{https://arxiv.org/abs/2101.08565}{{\ttfamily
  2101.08565}}.

\bibitem{Yennapureddy:2017vvb}
M.~K. Yennapureddy and F.~Melia, \emph{{Reconstruction of the HII Galaxy Hubble
  Diagram using Gaussian Processes}},
  \href{https://doi.org/10.1088/1475-7516/2017/11/029}{\emph{JCAP} {\bfseries
  11} (2017) 029} [\href{https://arxiv.org/abs/1711.03454}{{\ttfamily
  1711.03454}}].

\bibitem{Li:2019nux}
E.-K. Li, M.~Du, Z.-H. Zhou, H.~Zhang and L.~Xu, \emph{{Test the effects of
  $H_0$ on $f\sigma_8$ tension with Gaussian Process method}},
  \href{https://arxiv.org/abs/1911.12076}{{\ttfamily 1911.12076}}.

\bibitem{Seikel:2013fda}
M.~Seikel and C.~Clarkson, \emph{{Optimising Gaussian processes for
  reconstructing dark energy dynamics from supernovae}},
  \href{https://arxiv.org/abs/1311.6678}{{\ttfamily 1311.6678}}.

\bibitem{Seikel2012}
M.~Seikel, C.~Clarkson and M.~Smith, \emph{Reconstruction of dark energy and
  expansion dynamics using gaussian processes},
  \href{https://doi.org/10.1088/1475-7516/2012/06/036}{\emph{Journal of
  Cosmology and Astroparticle Physics} {\bfseries 2012} (2012) 036–036}.

\bibitem{Benisty:2020kdt}
D.~Benisty, \emph{{Quantifying the $S_8$ tension with the Redshift Space
  Distortion data set}},
  \href{https://doi.org/10.1016/j.dark.2020.100766}{\emph{Phys. Dark Univ.}
  {\bfseries 31} (2021) 100766}
  [\href{https://arxiv.org/abs/2005.03751}{{\ttfamily 2005.03751}}].

\bibitem{Belgacem:2019zzu}
E.~Belgacem, S.~Foffa, M.~Maggiore and T.~Yang, \emph{{Gaussian processes
  reconstruction of modified gravitational wave propagation}},
  \href{https://doi.org/10.1103/PhysRevD.101.063505}{\emph{Phys. Rev. D}
  {\bfseries 101} (2020) 063505}
  [\href{https://arxiv.org/abs/1911.11497}{{\ttfamily 1911.11497}}].

\bibitem{Moore:2015sza}
C.~J. Moore, C.~P.~L. Berry, A.~J.~K. Chua and J.~R. Gair, \emph{{Improving
  gravitational-wave parameter estimation using Gaussian process regression}},
  \href{https://doi.org/10.1103/PhysRevD.93.064001}{\emph{Phys. Rev. D}
  {\bfseries 93} (2016) 064001}
  [\href{https://arxiv.org/abs/1509.04066}{{\ttfamily 1509.04066}}].

\bibitem{Canas-Herrera:2021qxs}
G.~Ca\~nas Herrera, O.~Contigiani and V.~Vardanyan, \emph{{Learning how to
  surf: Reconstructing the propagation and origin of gravitational waves with
  Gaussian Processes}},  \href{https://arxiv.org/abs/2105.04262}{{\ttfamily
  2105.04262}}.

\bibitem{Reyes:2021owe}
M.~Reyes and C.~Escamilla-Rivera, \emph{{Improving data-driven
  model-independent reconstructions and new constraints in Horndeski
  cosmology}},  \href{https://arxiv.org/abs/2104.04484}{{\ttfamily
  2104.04484}}.

\bibitem{Briffa:2020qli}
R.~Briffa, S.~Capozziello, J.~Levi~Said, J.~Mifsud and E.~N. Saridakis,
  \emph{{Constraining teleparallel gravity through Gaussian processes}},
  \href{https://doi.org/10.1088/1361-6382/abd4f5}{\emph{Class. Quant. Grav.}
  {\bfseries 38} (2020) 055007}
  [\href{https://arxiv.org/abs/2009.14582}{{\ttfamily 2009.14582}}].

\bibitem{Cai:2019bdh}
Y.-F. Cai, M.~Khurshudyan and E.~N. Saridakis, \emph{{Model-independent
  reconstruction of $f(T)$ gravity from Gaussian Processes}},
  \href{https://doi.org/10.3847/1538-4357/ab5a7f}{\emph{Astrophys. J.}
  {\bfseries 888} (2020) 62}
  [\href{https://arxiv.org/abs/1907.10813}{{\ttfamily 1907.10813}}].

\bibitem{LeviSaid:2021yat}
J.~Levi~Said, J.~Mifsud, J.~Sultana and K.~Z. Adami, \emph{{Reconstructing
  teleparallel gravity with cosmic structure growth and expansion rate data}},
  \href{https://arxiv.org/abs/2103.05021}{{\ttfamily 2103.05021}}.

\bibitem{Cai:2015zoa}
T.~Yang, Z.-K. Guo and R.-G. Cai, \emph{{Reconstructing the interaction between
  dark energy and dark matter using Gaussian Processes}},
  \href{https://doi.org/10.1103/PhysRevD.91.123533}{\emph{Phys. Rev. D}
  {\bfseries 91} (2015) 123533}
  [\href{https://arxiv.org/abs/1505.04443}{{\ttfamily 1505.04443}}].

\bibitem{Montiel:2014fpa}
A.~Montiel, R.~Lazkoz, I.~Sendra, C.~Escamilla-Rivera and V.~Salzano,
  \emph{{Nonparametric reconstruction of the cosmic expansion with local
  regression smoothing and simulation extrapolation}},
  \href{https://doi.org/10.1103/PhysRevD.89.043007}{\emph{Phys. Rev. D}
  {\bfseries 89} (2014) 043007}
  [\href{https://arxiv.org/abs/1401.4188}{{\ttfamily 1401.4188}}].

\bibitem{Escamilla-Rivera:2015odt}
C.~Escamilla-Rivera and J.~C. Fabris, \emph{{Nonparametric reconstruction of
  the O$_m$ diagnostic to test $\Lambda$CDM}},
  \href{https://doi.org/10.3390/galaxies4040076}{\emph{Galaxies} {\bfseries 4}
  (2016) 76} [\href{https://arxiv.org/abs/1511.07066}{{\ttfamily 1511.07066}}].

\bibitem{Fernandez-Hernandez:2018yao}
L.~M. Fern\'andez-Hern\'andez, A.~Montiel and M.~A. Rodr\'\i{}guez-Meza,
  \emph{{Galaxy rotation curves using a non-parametric regression method: core,
  cuspy and fuzzy scalar field dark matter models}},
  \href{https://doi.org/10.1093/mnras/stz1969}{\emph{Mon. Not. Roy. Astron.
  Soc.} {\bfseries 488} (2019) 5127}
  [\href{https://arxiv.org/abs/1809.06875}{{\ttfamily 1809.06875}}].

\bibitem{Huterer:2002hy}
D.~Huterer and G.~Starkman, \emph{{Parameterization of dark-energy properties:
  A Principal-component approach}},
  \href{https://doi.org/10.1103/PhysRevLett.90.031301}{\emph{Phys. Rev. Lett.}
  {\bfseries 90} (2003) 031301}
  [\href{https://arxiv.org/abs/astro-ph/0207517}{{\ttfamily
  astro-ph/0207517}}].

\bibitem{Espana-Bonet:2005wkl}
C.~Espana-Bonet and P.~Ruiz-Lapuente, \emph{{Dark energy as an inverse
  problem}},  \href{https://arxiv.org/abs/hep-ph/0503210}{{\ttfamily
  hep-ph/0503210}}.

\bibitem{Bonvin:2006en}
C.~Bonvin, R.~Durrer and M.~Kunz, \emph{{The dipole of the luminosity distance:
  a direct measure of H(z)}},
  \href{https://doi.org/10.1103/PhysRevLett.96.191302}{\emph{Phys. Rev. Lett.}
  {\bfseries 96} (2006) 191302}
  [\href{https://arxiv.org/abs/astro-ph/0603240}{{\ttfamily
  astro-ph/0603240}}].

\bibitem{AlbertoVazquez:2012ofj}
J.~Alberto~Vazquez, M.~Bridges, M.~P. Hobson and A.~N. Lasenby,
  \emph{{Reconstruction of the Dark Energy equation of state}},
  \href{https://doi.org/10.1088/1475-7516/2012/09/020}{\emph{JCAP} {\bfseries
  09} (2012) 020} [\href{https://arxiv.org/abs/1205.0847}{{\ttfamily
  1205.0847}}].

\bibitem{Bogdanos_2009}
C.~Bogdanos and S.~Nesseris, \emph{Genetic algorithms and supernovae type ia
  analysis}, \href{https://doi.org/10.1088/1475-7516/2009/05/006}{\emph{Journal
  of Cosmology and Astroparticle Physics} {\bfseries 2009} (2009) 006–006}.

\bibitem{Daly_2003}
R.~A. Daly and S.~G. Djorgovski, \emph{A model‐independent determination of
  the expansion and acceleration rates of the universe as a function of
  redshift and constraints on dark energy},
  \href{https://doi.org/10.1086/378230}{\emph{The Astrophysical Journal}
  {\bfseries 597} (2003) 9–20}.

\bibitem{Fay_2006}
S.~Fay and R.~Tavakol, \emph{Model-independent dark energy reconstruction
  scheme using the geometrical form of the luminosity-distance relation},
  \href{https://doi.org/10.1103/physrevd.74.083513}{\emph{Physical Review D}
  {\bfseries 74} (2006) }.

\bibitem{Benitez_Herrera_2011}
S.~Benitez-Herrera, F.~Röpke, W.~Hillebrandt, C.~Mignone, M.~Bartelmann and
  J.~Weller, \emph{Model-independent reconstruction of the expansion history of
  the universe from type ia supernovae},
  \href{https://doi.org/10.1111/j.1365-2966.2011.19716.x}{\emph{Monthly Notices
  of the Royal Astronomical Society} {\bfseries 419} (2011) 513–521}.

\bibitem{Holsclaw_2010}
T.~Holsclaw, U.~Alam, B.~Sansó, H.~Lee, K.~Heitmann, S.~Habib et~al.,
  \emph{Nonparametric dark energy reconstruction from supernova data},
  \href{https://doi.org/10.1103/physrevlett.105.241302}{\emph{Physical Review
  Letters} {\bfseries 105} (2010) }.

\bibitem{Montiel_2014}
A.~Montiel, R.~Lazkoz, I.~Sendra, C.~Escamilla-Rivera and V.~Salzano,
  \emph{Nonparametric reconstruction of the cosmic expansion with local
  regression smoothing and simulation extrapolation},
  \href{https://doi.org/10.1103/physrevd.89.043007}{\emph{Physical Review D}
  {\bfseries 89} (2014) }.

\bibitem{Bernardo:2021mfs}
R.~C. Bernardo and J.~Levi~Said, \emph{{Towards a model-independent
  reconstruction approach for late-time Hubble data}},
  \href{https://arxiv.org/abs/2106.08688}{{\ttfamily 2106.08688}}.

\bibitem{Guo:2018ans}
R.-Y. Guo, J.-F. Zhang and X.~Zhang, \emph{{Can the $H_0$ tension be resolved
  in extensions to $\Lambda$CDM cosmology?}},
  \href{https://doi.org/10.1088/1475-7516/2019/02/054}{\emph{JCAP} {\bfseries
  02} (2019) 054} [\href{https://arxiv.org/abs/1809.02340}{{\ttfamily
  1809.02340}}].

\bibitem{Lambiase:2018ows}
G.~Lambiase, S.~Mohanty, A.~Narang and P.~Parashari, \emph{{Testing dark energy
  models in the light of $\sigma _8$ tension}},
  \href{https://doi.org/10.1140/epjc/s10052-019-6634-6}{\emph{Eur. Phys. J. C}
  {\bfseries 79} (2019) 141}
  [\href{https://arxiv.org/abs/1804.07154}{{\ttfamily 1804.07154}}].

\bibitem{Rani:2015lia}
N.~Rani, D.~Jain, S.~Mahajan, A.~Mukherjee and N.~Pires, \emph{{Transition
  Redshift: New constraints from parametric and nonparametric methods}},
  \href{https://doi.org/10.1088/1475-7516/2015/12/045}{\emph{JCAP} {\bfseries
  12} (2015) 045} [\href{https://arxiv.org/abs/1503.08543}{{\ttfamily
  1503.08543}}].

\bibitem{10.5555/971143}
D.~J.~C. MacKay, \emph{Information Theory, Inference \& Learning Algorithms}.
  Cambridge University Press, USA, 2002.

\bibitem{Busti:2014aoa}
V.~C. Busti, C.~Clarkson and M.~Seikel, \emph{{The Value of $H_0$ from Gaussian
  Processes}}, \href{https://doi.org/10.1017/S1743921314013751}{\emph{IAU
  Symp.} {\bfseries 306} (2014) 25}
  [\href{https://arxiv.org/abs/1407.5227}{{\ttfamily 1407.5227}}].

\bibitem{Busti:2014dua}
V.~C. Busti, C.~Clarkson and M.~Seikel, \emph{{Evidence for a Lower Value for
  $H_0$ from Cosmic Chronometers Data?}},
  \href{https://doi.org/10.1093/mnrasl/slu035}{\emph{Mon. Not. Roy. Astron.
  Soc.} {\bfseries 441} (2014) 11}
  [\href{https://arxiv.org/abs/1402.5429}{{\ttfamily 1402.5429}}].

\bibitem{Verde:2014qea}
L.~Verde, P.~Protopapas and R.~Jimenez, \emph{{The expansion rate of the
  intermediate Universe in light of Planck}},
  \href{https://doi.org/10.1016/j.dark.2014.09.003}{\emph{Phys. Dark Univ.}
  {\bfseries 5-6} (2014) 307}
  [\href{https://arxiv.org/abs/1403.2181}{{\ttfamily 1403.2181}}].

\bibitem{Li:2015nta}
Z.~Li, J.~E. Gonzalez, H.~Yu, Z.-H. Zhu and J.~S. Alcaniz, \emph{{Constructing
  a cosmological model-independent Hubble diagram of type Ia supernovae with
  cosmic chronometers}},
  \href{https://doi.org/10.1103/PhysRevD.93.043014}{\emph{Phys. Rev. D}
  {\bfseries 93} (2016) 043014}
  [\href{https://arxiv.org/abs/1504.03269}{{\ttfamily 1504.03269}}].

\bibitem{Shafieloo:2012ht}
A.~Shafieloo, A.~G. Kim and E.~V. Linder, \emph{{Gaussian Process
  Cosmography}}, \href{https://doi.org/10.1103/PhysRevD.85.123530}{\emph{Phys.
  Rev. D} {\bfseries 85} (2012) 123530}
  [\href{https://arxiv.org/abs/1204.2272}{{\ttfamily 1204.2272}}].

\bibitem{Cai:2015pia}
R.-G. Cai, Z.-K. Guo and T.~Yang, \emph{{Null test of the cosmic curvature
  using $H(z)$ and supernovae data}},
  \href{https://doi.org/10.1103/PhysRevD.93.043517}{\emph{Phys. Rev. D}
  {\bfseries 93} (2016) 043517}
  [\href{https://arxiv.org/abs/1509.06283}{{\ttfamily 1509.06283}}].

\bibitem{Wang:2017jdm}
D.~Wang and X.-H. Meng, \emph{{Improved constraints on the dark energy equation
  of state using Gaussian processes}},
  \href{https://doi.org/10.1103/PhysRevD.95.023508}{\emph{Phys. Rev. D}
  {\bfseries 95} (2017) 023508}
  [\href{https://arxiv.org/abs/1708.07750}{{\ttfamily 1708.07750}}].

\bibitem{Zhou:2019gda}
L.~Zhou, X.~Fu, Z.~Peng and J.~Chen, \emph{{Probing the Cosmic Opacity from
  Future Gravitational Wave Standard Sirens}},
  \href{https://doi.org/10.1103/PhysRevD.100.123539}{\emph{Phys. Rev. D}
  {\bfseries 100} (2019) 123539}
  [\href{https://arxiv.org/abs/1912.02327}{{\ttfamily 1912.02327}}].

\bibitem{Mukherjee:2020vkx}
P.~Mukherjee and N.~Banerjee, \emph{{Revisiting a non-parametric reconstruction
  of the deceleration parameter from observational data}},
  \href{https://arxiv.org/abs/2007.15941}{{\ttfamily 2007.15941}}.

\bibitem{Zhang:2018gjb}
M.-J. Zhang and H.~Li, \emph{{Gaussian processes reconstruction of dark energy
  from observational data}},
  \href{https://doi.org/10.1140/epjc/s10052-018-5953-3}{\emph{Eur. Phys. J. C}
  {\bfseries 78} (2018) 460}
  [\href{https://arxiv.org/abs/1806.02981}{{\ttfamily 1806.02981}}].

\bibitem{Aljaf:2020eqh}
M.~Aljaf, D.~Gregoris and M.~Khurshudyan, \emph{{Constraints on interacting
  dark energy models through cosmic chronometers and Gaussian process}},
  \href{https://arxiv.org/abs/2005.01891}{{\ttfamily 2005.01891}}.

\bibitem{Liao:2019qoc}
K.~Liao, A.~Shafieloo, R.~E. Keeley and E.~V. Linder, \emph{{A
  model-independent determination of the Hubble constant from lensed quasars
  and supernovae using Gaussian process regression}},
  \href{https://doi.org/10.3847/2041-8213/ab5308}{\emph{Astrophys. J. Lett.}
  {\bfseries 886} (2019) L23}
  [\href{https://arxiv.org/abs/1908.04967}{{\ttfamily 1908.04967}}].

\bibitem{Yu:2017iju}
H.~Yu, B.~Ratra and F.-Y. Wang, \emph{{Hubble Parameter and Baryon Acoustic
  Oscillation Measurement Constraints on the Hubble Constant, the Deviation
  from the Spatially Flat \ensuremath{\Lambda}CDM Model, the
  Deceleration\textendash{}Acceleration Transition Redshift, and Spatial
  Curvature}}, \href{https://doi.org/10.3847/1538-4357/aab0a2}{\emph{Astrophys.
  J.} {\bfseries 856} (2018) 3}
  [\href{https://arxiv.org/abs/1711.03437}{{\ttfamily 1711.03437}}].

\bibitem{Krishnan:2020vaf}
C.~Krishnan, E.~Colg\'ain, M.~M. Sheikh-Jabbari and T.~Yang, \emph{{Running
  Hubble Tension and a H0 Diagnostic}},
  \href{https://arxiv.org/abs/2011.02858}{{\ttfamily 2011.02858}}.

\bibitem{Moresco:2015cya}
M.~Moresco, \emph{{Raising the bar: new constraints on the Hubble parameter
  with cosmic chronometers at z \ensuremath{\sim} 2}},
  \href{https://doi.org/10.1093/mnrasl/slv037}{\emph{Mon. Not. Roy. Astron.
  Soc.} {\bfseries 450} (2015) L16}
  [\href{https://arxiv.org/abs/1503.01116}{{\ttfamily 1503.01116}}].

\bibitem{Moresco:2016mzx}
M.~Moresco et~al., \emph{{A 6\% measurement of the Hubble parameter at
  $z\sim0.45$: direct evidence of the epoch of cosmic re-acceleration}},
  \href{https://doi.org/10.1088/1475-7516/2016/05/014}{\emph{JCAP} {\bfseries
  05} (2016) 014} [\href{https://arxiv.org/abs/1601.01701}{{\ttfamily
  1601.01701}}].

\bibitem{2010JCAP...02..008S}
D.~{Stern}, R.~{Jimenez}, L.~{Verde}, M.~{Kamionkowski} and S.~A. {Stanford},
  \emph{{Cosmic chronometers: constraining the equation of state of dark
  energy. I: H(z) measurements}},
  \href{https://doi.org/10.1088/1475-7516/2010/02/008}{\emph{JCAP} {\bfseries
  2010} (2010) 008} [\href{https://arxiv.org/abs/0907.3149}{{\ttfamily
  0907.3149}}].

\bibitem{2012JCAP...08..006M}
M.~{Moresco}, A.~{Cimatti}, R.~{Jimenez}, L.~{Pozzetti}, G.~{Zamorani},
  M.~{Bolzonella} et~al., \emph{{Improved constraints on the expansion rate of
  the Universe up to z \raisebox{-0.5ex}\textasciitilde 1.1 from the
  spectroscopic evolution of cosmic chronometers}},
  \href{https://doi.org/10.1088/1475-7516/2012/08/006}{\emph{JCAP} {\bfseries
  2012} (2012) 006} [\href{https://arxiv.org/abs/1201.3609}{{\ttfamily
  1201.3609}}].

\bibitem{2014RAA....14.1221Z}
C.~{Zhang}, H.~{Zhang}, S.~{Yuan}, S.~{Liu}, T.-J. {Zhang} and Y.-C. {Sun},
  \emph{{Four new observational H(z) data from luminous red galaxies in the
  Sloan Digital Sky Survey data release seven}},
  \href{https://doi.org/10.1088/1674-4527/14/10/002}{\emph{Research in
  Astronomy and Astrophysics} {\bfseries 14} (2014) 1221}
  [\href{https://arxiv.org/abs/1207.4541}{{\ttfamily 1207.4541}}].

\bibitem{BOSS:2013igd}
{\scshape BOSS} collaboration, \emph{{Quasar-Lyman $\alpha$ Forest
  Cross-Correlation from BOSS DR11 : Baryon Acoustic Oscillations}},
  \href{https://doi.org/10.1088/1475-7516/2014/05/027}{\emph{JCAP} {\bfseries
  05} (2014) 027} [\href{https://arxiv.org/abs/1311.1767}{{\ttfamily
  1311.1767}}].

\bibitem{BOSS:2014hwf}
{\scshape BOSS} collaboration, \emph{{Baryon acoustic oscillations in the
  Ly\ensuremath{\alpha} forest of BOSS DR11 quasars}},
  \href{https://doi.org/10.1051/0004-6361/201423969}{\emph{Astron. Astrophys.}
  {\bfseries 574} (2015) A59}
  [\href{https://arxiv.org/abs/1404.1801}{{\ttfamily 1404.1801}}].

\bibitem{Bautista:2017zgn}
J.~E. Bautista et~al., \emph{{Measurement of baryon acoustic oscillation
  correlations at $z=2.3$ with SDSS DR12 Ly$\alpha$-Forests}},
  \href{https://doi.org/10.1051/0004-6361/201730533}{\emph{Astron. Astrophys.}
  {\bfseries 603} (2017) A12}
  [\href{https://arxiv.org/abs/1702.00176}{{\ttfamily 1702.00176}}].

\bibitem{BOSS:2013rlg}
{\scshape BOSS} collaboration, \emph{{The clustering of galaxies in the
  SDSS-III Baryon Oscillation Spectroscopic Survey: baryon acoustic
  oscillations in the Data Releases 10 and 11 Galaxy samples}},
  \href{https://doi.org/10.1093/mnras/stu523}{\emph{Mon. Not. Roy. Astron.
  Soc.} {\bfseries 441} (2014) 24}
  [\href{https://arxiv.org/abs/1312.4877}{{\ttfamily 1312.4877}}].

\bibitem{2012MNRAS.425..405B}
C.~{Blake}, S.~{Brough}, M.~{Colless}, C.~{Contreras}, W.~{Couch}, S.~{Croom}
  et~al., \emph{{The WiggleZ Dark Energy Survey: joint measurements of the
  expansion and growth history at z < 1}},
  \href{https://doi.org/10.1111/j.1365-2966.2012.21473.x}{\emph{Mon. Not. Roy.
  Astron. Soc.} {\bfseries 425} (2012) 405}
  [\href{https://arxiv.org/abs/1204.3674}{{\ttfamily 1204.3674}}].

\bibitem{2017MNRAS.470.2617A}
S.~{Alam}, M.~{Ata}, S.~{Bailey}, F.~{Beutler}, D.~{Bizyaev}, J.~A. {Blazek}
  et~al., \emph{{The clustering of galaxies in the completed SDSS-III Baryon
  Oscillation Spectroscopic Survey: cosmological analysis of the DR12 galaxy
  sample}}, \href{https://doi.org/10.1093/mnras/stx721}{\emph{Mon. Not. Roy.
  Astron. Soc.} {\bfseries 470} (2017) 2617}
  [\href{https://arxiv.org/abs/1607.03155}{{\ttfamily 1607.03155}}].

\bibitem{10.1093/mnras/stt1290}
C.-H. Chuang and Y.~Wang, \emph{{Modelling the anisotropic two-point galaxy
  correlation function on small scales and single-probe measurements of H(z),
  DA(z) and $f(z)\sigma_8 (z)$ from the Sloan Digital Sky Survey DR7 luminous
  red galaxies}}, \href{https://doi.org/10.1093/mnras/stt1290}{\emph{Mon. Not.
  Roy. Astron. Soc.} {\bfseries 435} (2013) 255}
  [\href{https://arxiv.org/abs/https://academic.oup.com/mnras/article-pdf/435/1/255/3865066/stt1290.pdf}{{\ttfamily
  https://academic.oup.com/mnras/article-pdf/435/1/255/3865066/stt1290.pdf}}].

\bibitem{BOSS:2016zkm}
{\scshape BOSS} collaboration, \emph{{The clustering of galaxies in the
  completed SDSS-III Baryon Oscillation Spectroscopic Survey: tomographic BAO
  analysis of DR12 combined sample in configuration space}},
  \href{https://doi.org/10.1093/mnras/stx1090}{\emph{Mon. Not. Roy. Astron.
  Soc.} {\bfseries 469} (2017) 3762}
  [\href{https://arxiv.org/abs/1607.03154}{{\ttfamily 1607.03154}}].

\bibitem{Oka:2013cba}
A.~Oka, S.~Saito, T.~Nishimichi, A.~Taruya and K.~Yamamoto, \emph{{Simultaneous
  constraints on the growth of structure and cosmic expansion from the
  multipole power spectra of the SDSS DR7 LRG sample}},
  \href{https://doi.org/10.1093/mnras/stu111}{\emph{Mon. Not. Roy. Astron.
  Soc.} {\bfseries 439} (2014) 2515}
  [\href{https://arxiv.org/abs/1310.2820}{{\ttfamily 1310.2820}}].

\bibitem{Gaztanaga:2008xz}
E.~Gaztanaga, A.~Cabre and L.~Hui, \emph{{Clustering of Luminous Red Galaxies
  IV: Baryon Acoustic Peak in the Line-of-Sight Direction and a Direct
  Measurement of H(z)}},
  \href{https://doi.org/10.1111/j.1365-2966.2009.15405.x}{\emph{Mon. Not. Roy.
  Astron. Soc.} {\bfseries 399} (2009) 1663}
  [\href{https://arxiv.org/abs/0807.3551}{{\ttfamily 0807.3551}}].

\bibitem{Magana:2017nfs}
J.~Magana, M.~H. Amante, M.~A. Garcia-Aspeitia and V.~Motta, \emph{{The
  Cardassian expansion revisited: constraints from updated Hubble parameter
  measurements and type Ia supernova data}},
  \href{https://doi.org/10.1093/mnras/sty260}{\emph{Mon. Not. Roy. Astron.
  Soc.} {\bfseries 476} (2018) 1036}
  [\href{https://arxiv.org/abs/1706.09848}{{\ttfamily 1706.09848}}].

\bibitem{Nunes:2020hzy}
R.~C. Nunes, S.~K. Yadav, J.~F. Jesus and A.~Bernui, \emph{{Cosmological
  parameter analyses using transversal BAO data}},
  \href{https://doi.org/10.1093/mnras/staa2036}{\emph{Mon. Not. Roy. Astron.
  Soc.} {\bfseries 497} (2020) 2133}
  [\href{https://arxiv.org/abs/2002.09293}{{\ttfamily 2002.09293}}].

\bibitem{Scolnic:2017caz}
D.~M. Scolnic et~al., \emph{{The Complete Light-curve Sample of
  Spectroscopically Confirmed SNe Ia from Pan-STARRS1 and Cosmological
  Constraints from the Combined Pantheon Sample}},
  \href{https://doi.org/10.3847/1538-4357/aab9bb}{\emph{Astrophys. J.}
  {\bfseries 859} (2018) 101}
  [\href{https://arxiv.org/abs/1710.00845}{{\ttfamily 1710.00845}}].

\bibitem{freedman2019carnegie}
W.~L. Freedman, B.~F. Madore, D.~Hatt, T.~J. Hoyt, I.~S. Jang, R.~L. Beaton
  et~al., \emph{{The Carnegie-Chicago Hubble Program. VIII. An independent
  determination of the Hubble constant based on the tip of the red giant
  branch}}, {\emph{The Astrophysical Journal} {\bfseries 882} (2019) 34}.

\bibitem{wong2020h0licow}
K.~C. Wong, S.~H. Suyu, G.~C. Chen, C.~E. Rusu, M.~Millon, D.~Sluse et~al.,
  \emph{{H0LiCOW--XIII. A 2.4 percent measurement of H0 from lensed quasars:
  5.3 $\sigma$ tension between early-and late-Universe probes}}, {\emph{Monthly
  Notices of the Royal Astronomical Society} {\bfseries 498} (2020) 1420}.

\bibitem{Riess:2016jrr}
A.~G. Riess et~al., \emph{{{A 2.4\% Determination of the Local Value of the
  Hubble Constant}}},
  \href{https://doi.org/10.3847/0004-637X/826/1/56}{\emph{Astrophys. J.}
  {\bfseries 826} (2016) 56}
  [\href{https://arxiv.org/abs/1604.01424}{{\ttfamily 1604.01424}}].

\bibitem{Chiang:2017yrq}
H.~W. Chiang, A.~E. Romano, F.~Nugier and P.~Chen, \emph{{Probing homogeneity
  with standard candles}},
  \href{https://doi.org/10.1088/1475-7516/2019/11/016}{\emph{JCAP} {\bfseries
  11} (2019) 016} [\href{https://arxiv.org/abs/1706.09734}{{\ttfamily
  1706.09734}}].

\bibitem{Di_Valentino_2021}
E.~Di~Valentino, O.~Mena, S.~Pan, L.~Visinelli, W.~Yang, A.~Melchiorri et~al.,
  \emph{In the realm of the hubble tension—a review of solutions *},
  \href{https://doi.org/10.1088/1361-6382/ac086d}{\emph{Classical and Quantum
  Gravity} {\bfseries 38} (2021) 153001}.

\bibitem{DiValentino:2020zio}
E.~Di~Valentino et~al., \emph{{Snowmass2021 - Letter of interest cosmology
  intertwined II: The hubble constant tension}},
  \href{https://doi.org/10.1016/j.astropartphys.2021.102605}{\emph{Astropart.
  Phys.} {\bfseries 131} (2021) 102605}
  [\href{https://arxiv.org/abs/2008.11284}{{\ttfamily 2008.11284}}].

\bibitem{Freedman:2021ahq}
W.~L. Freedman, \emph{{Measurements of the Hubble Constant: Tensions in
  Perspective}},  \href{https://arxiv.org/abs/2106.15656}{{\ttfamily
  2106.15656}}.

\bibitem{Alestas:2021xes}
G.~Alestas and L.~Perivolaropoulos, \emph{{Late-time approaches to the Hubble
  tension deforming H(z), worsen the growth tension}},
  \href{https://doi.org/10.1093/mnras/stab1070}{\emph{Mon. Not. Roy. Astron.
  Soc.} {\bfseries 504} (2021) 3956}
  [\href{https://arxiv.org/abs/2103.04045}{{\ttfamily 2103.04045}}].

\end{thebibliography}\endgroup

\end{document}